%% file: usenix.tex
\documentclass[letterpaper,twocolumn,10pt]{article}
\usepackage{usenix,epsfig}

\usepackage{colortbl, xcolor}
\usepackage[utf8]{inputenc}
\usepackage{CJKutf8}
\usepackage{array,booktabs,tabularx}

\usepackage{booktabs}
\usepackage{listings}

\lstdefinestyle{promptstyle}{
    basicstyle=\ttfamily\small, %
    breaklines=true,            %
    breakatwhitespace=true,     %
    postbreak=\mbox{\textcolor{red}{$\hookrightarrow$}\space}, %
    frame=single,               %
    framerule=0pt,              %
    framesep=10pt,              %
    backgroundcolor=\color{black!5} %
}

\usepackage{xspace} %
\usepackage{graphicx}
\usepackage{bm}
\usepackage{xurl}
\newcommand{\sysname}{\textsc{Chatterbox}\xspace}

\newcommand{\hllm}{\textsc{HLLM}\xspace}
\newcommand{\attackers}{\textsc{scammers}\xspace}

\begin{document}

\date{}

\title{\Large \bf Victim as a Service: Designing a System for Engaging with Interactive Scammers}

\author{
{\rm Daniel Spokoyny\textsuperscript{*}}
\and
{\rm Nikolai Vogler\textsuperscript{*}}
\and
{\rm Xin Gao\textsuperscript{+}}
\and
{\rm Tianyi Zheng\textsuperscript{+}}
\and
{\rm Yufei Weng}
\and
{\rm Jonghyun Park}
\and
{\rm Jiajun Jiao}
\and
{\rm Geoffrey M. Voelker}
\and
{\rm Stefan Savage}
\and
{\rm Taylor Berg-Kirkpatrick}
\\[1.5ex]
{UC San Diego}
} %

\maketitle

{
\renewcommand{\thefootnote}{}
\footnotetext{\textsuperscript{*}, \textsuperscript{+}: Denotes equal contribution.}
}

\thispagestyle{empty}

\subsection*{Abstract}
\input{abstract}
\input{intro}

\begin{figure*}[t]
\centering
\includegraphics[width=0.82\textwidth]{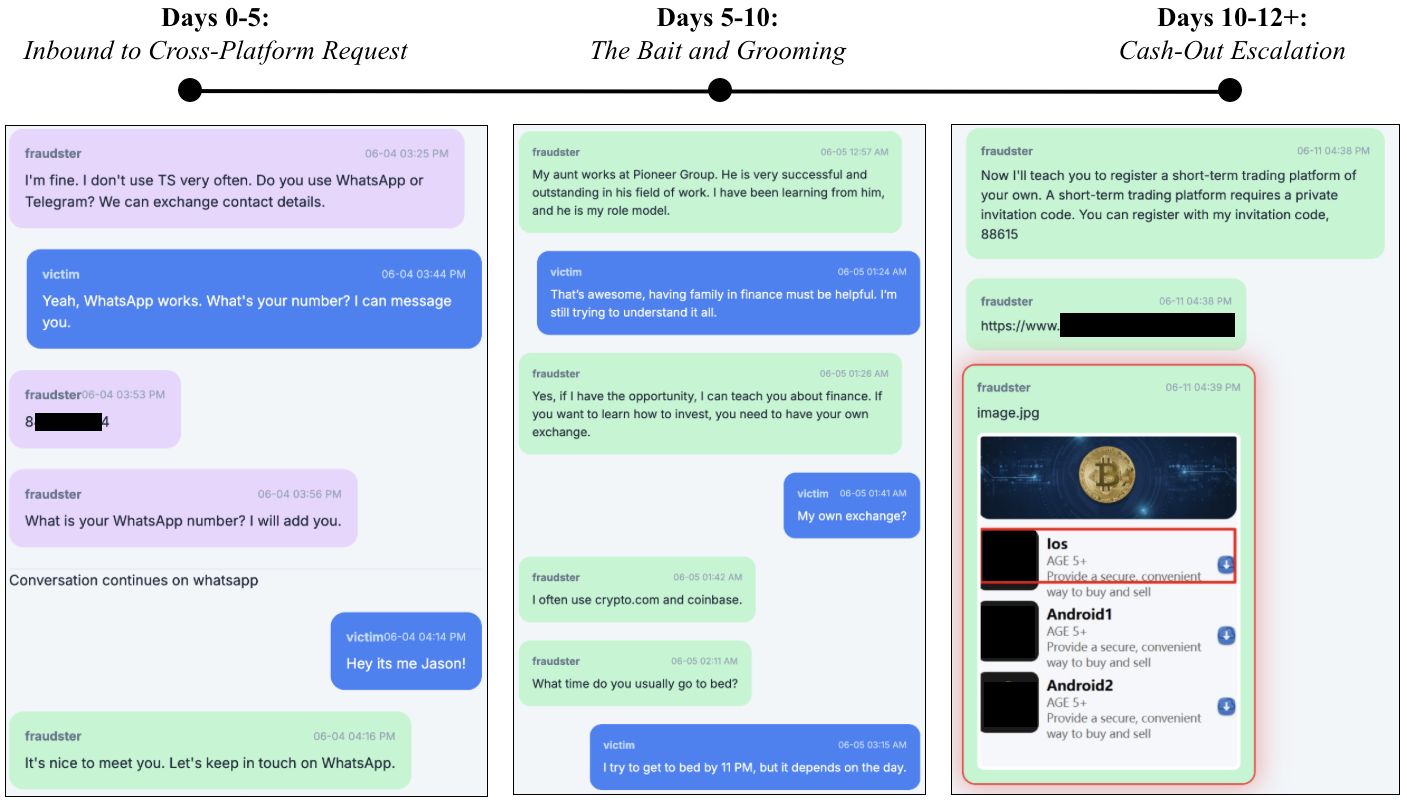}
\caption{In this example of a scam attempt against \sysname, we illustrate three key phases: (left) the \emph{cross-platform} request, which occurs 5 days after initial contact, (center) casual introduction to \emph{the bait}, which is an investment opportunity, and (right) the cash-out phase where the scammer attempts to extract financial value using a fraudulent app/website. 
Each phase presents unique challenges and requirements for our system to effectively mimic human behavior and maintain believability. 
See Section~\ref{sec:motivation} for a detailed walkthrough of this example and the system requirements it motivates.}
\label{fig:example_walkthrough}
\end{figure*}
\input{background}

\begin{figure}[t]
    \centering
    \includegraphics[width=\columnwidth]{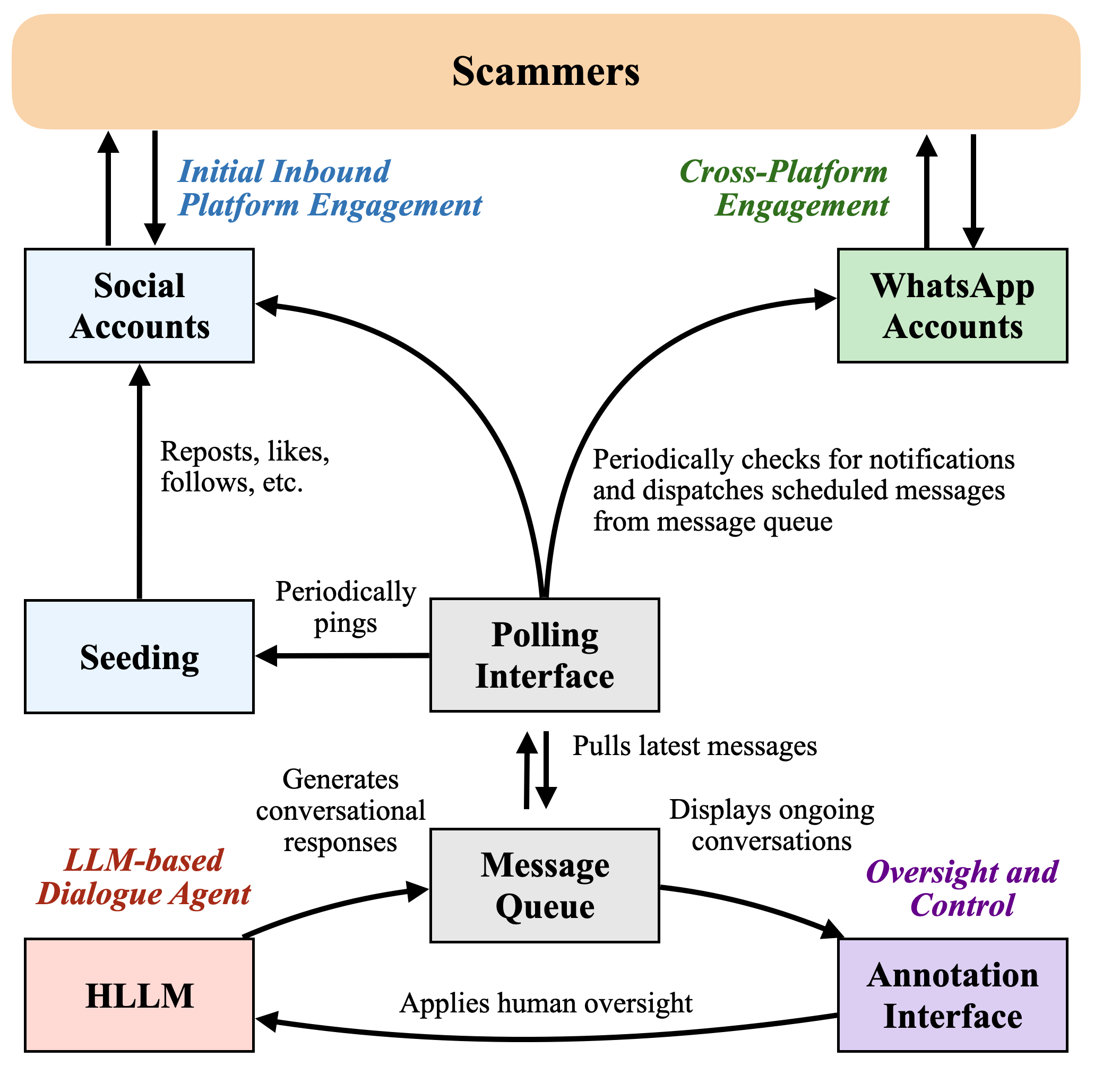}
    \caption{System Architecture of \sysname. The system attracts initial inbound engagement from scammers by using a seeding module to generate realistic activity (e.g., reposts, likes) on social media accounts (Sec~\ref{sec:system_seeding}). A polling module periodically retrieves incoming messages, which are placed into a message queue. From the queue, messages are sent to the HLLM (Sec~\ref{sec:honeypot_llm}) to generate conversational responses and to a human-in-the-loop annotation interface for oversight and control (Sec~\ref{sec:annotate}). The architecture supports the entire engagement lifecycle, including cross-platform migration to WhatsApp accounts to continue the conversation (Sec~\ref{sec:system_cross_platform})}
    \label{fig:system_architecture}
\end{figure}
\input{motivation}
\input{honeypot_llm}
\input{system}

\input{experience}

\input{limitations_future}

\appendix
\cleardoublepage
\input{ethics}

\input{appendix_persona_generation}

\input{appendix}
\input{appendix_prompts}
\input{appendix_honeypot_llm}

{\footnotesize \bibliographystyle{acm}
\bibliography{sample}}

\end{document}

%% file: abstract.tex
Pig butchering, and similar interactive online scams, lower their victims' defenses by building trust over extended periods of conversation -- sometimes weeks or months.  They have become increasingly common and the source of significant public losses (at least \$75B by one recent study).  However, because of their long-term conversational nature, they are extremely challenging to investigate at scale.   In this paper, we describe the motivation, design, implementation, and experience with \sysname, an LLM-based system that automates long-term engagement with online scammers, making large-scale investigations of their tactics possible. We describe the techniques we have developed to attract scam attempts, the system and LLM-engineering required to convincingly engage with scammers, and the necessary capabilities required to satisfy or evade ``milestones'' in scammers' workflow.  

%% file: intro.tex
\section{Introduction}
Long-term interactive trust building is a tactic of particular online scams that build a relationship with the victim over an extended time period (days, weeks, or even months) and then exploit the developed  rapport for some criminal purpose.  While such scams are both widespread and costly (e.g., one study of pig butchering profits between 2020--2024 placed losses at \$75B~\cite{griffin2024crypto}), it is challenging to gather data on the nature of the scams themselves, the techniques employed to manipulate users or the range of mechanisms used to extract value.  Few complete transcripts exist and undercover work requires long-term manual engagement in a manner that precludes data collection at scale.  

Our paper is motivated by this need and presents a new approach made possible by recent advances in Large Language Models (LLMs). While undercover work has historically required long-term manual engagement that precludes data collection at scale, the modern ability of LLMs to conduct convincing, open-ended conversation now makes automated investigation feasible. Harnessing this new capability, we describe the design and implementation of \sysname, a prototype system whose primary function is to automatically collect unsolicited scam attempts and successfully engage with these parties on an ongoing and \emph{controllable} basis.  In particular, we focus on text-based scams (i.e., solicited via text message, third-party messaging apps or online service direct message) in which the ultimate criminal goal is not immediately clear (i.e., we do not engage with ``urgent'' online scams, such as fake anti-virus, technical support, lottery, or IRS debt scams in which the ``ask'' is evident immediately).  Abstractly, \sysname seeds victim personas online, mediates any subsequent attempt at contact, and simulates its victim personas using a customized `Honeypot' Large Language Model (HLLM) for conversational language generation.

However, unlike honeypot systems for more traditional ``urgent''
scams, developing a honeypot for interactive online scams requires a
range of new capabilities to increase victim verisimilitude and meet
scammer requirements.  These capabilities include maintaining
conversational context across platforms (e.g., from Truth Social to
WhatsApp) and long time periods (e.g, days to weeks), multiplexing
limited resources (e.g., phones) among many simultaneous interactions,
absorbing and threading multi-modal content (e.g., photos and audio)
into conversations, creating rich victim personas to respond to a wide
range of social conversational prompts (e.g., about family, interests,
weather, sports) and challenges (e.g., contextual selfies as proof of
life), etc.  Moreover, when \sysname is unable to support a
particular request made by a scammer (e.g., transferring to a
messaging platform it does not yet support, such as Telegram) it
exploits the flexibility of its LLM (and the motivated nature of
scammers) to offer alternatives that are supported.

The goal of \sysname is to support large-scale data collection of representative attacker transcripts.  These in turn can be used to drive defenses (e.g., training client-side classifiers to identify scammer-like behavior) or affirmative interventions (e.g., identifying the apps and payment endpoints being used by different groups of scammers to support both attribution and financial ``takedowns''~\cite{Huang2018,lubbertsenghost,DOJ2025CryptoSeizure}).  Moreover, the flexibility inherent in the LLM-based nature of \sysname allows for controlled insertion of obstacles (e.g., ``Sadly, I don't have an Android phone... is there a way to make this work with an iPhone?'') or opportunities (e.g., ``My neighbor just paid for a new pool with his crypto investments.  I wish I knew how to do that.'') at scale, to support a range of analyses concerning attacker tactics and goals and their adaptability to different victim groups.

%% file: background.tex
\section{Background}

Electronic communications systems have long been used as vectors for fraud, as they reduce cost and friction for reaching ever larger numbers of potential victims.  The advent of the telegraph led directly to the creation of the criminal wire fraud statute, the introduction of e-mail, was swiftly followed by spam and phishing. So too social networks, SMS, and so on have all become fodder for efforts to defraud victims of their money or property. 

Roughly speaking, we can divide scamming activities into three classes, based on the length of interaction and its depth or sophistication in engaging with targets.  At the short, low-friction end of the spectrum are grifts built around one (or a few) high-urgency lures.  These include a wide array of standard scams including fake lottery/prize notification, advance-fee fraud, parcel-fee requests, etc).  Such lures can be sent en masse in an automated fashion and monetization can frequently occur online, absent much individualized labor from a scammer before payout~\cite{surveyscams,pumpdump,craigslist,fakeav,doublenothing23,charityscam}.   Medium-interaction scams, such as tech support or account recovery fraud, must sustain a conversational session long enough to obtain remote access, credentials, or an initial payment, sometimes pivoting platforms (email $\rightarrow$ phone $\rightarrow$ remote desktop) but rarely persisting beyond days.~\cite{dialoneforscam17,snorcall23,heymum25,conningconman24,scamchatbot24}  These scams also use automated lures focused on urgency, but then divert respondents to collections of on-call scammers to handle the short, but intense interactive phase of the effort.\footnote{Note that sometimes there may be multiple tiers of interactive scammers with one tier responsible for obtaining access, and then a second tier which handles payout.}  In general, both such categories of scams tend to be focused on high-volume efforts with modest per-victim return (i.e., < \$2k).

In pursuit of progressively larger financial returns and evasion of increasingly effective platform defenses (e.g., blacklists, content filters, etc. -- e.g.,~\cite{telblack18}) and user wariness, some scammers have shifted toward higher-effort approaches that deliberately cultivate trust over weeks or months across multiple channels and modalities to make victims feel they are willfully, even rationally, cooperating.  So-called ``pig butchering'' attacks (named to reflect the notion of ``fattening'', or cultivating, the victim before they are fully exploited) first emerged in China in early 2016, but have since spread world-wide.~\cite{sophos23,proofpoint23,wired24}  Unlike low-interaction scams, here scammers attempt to engage potential victims with low-urgency requests such as mistakenly addressed text messages or commenting on social media about some claimed shared interest.  Those who respond to such modest engagement (likely selecting for a more vulnerable population) then receive further interaction, building rapport slowly over time (frequently romantic in nature) -- just as a real inter-personal relationship might build.  Such scams can take weeks or months to build trust before a target is asked for money; indeed, one common approach is not to ask for money at all, but to entice the victim into asking the scammer if they may invest.  

However, the payoff from developing this trust is that, when successful, such attacks can relieve victims of large sums of money (with losses of tens or hundreds of thousands of dollars).  One recent study, based on cryptocurrency clustering of known pig butchering wallets, estimated direct losses from such attacks at \$75B between 2020 and 2024~\cite{griffin2024crypto}).  Sadly, those engaging with pig butchering victims are frequently victims themselves; there is an array of evidence indicating that this scamming labor pool is largely composed of forced labor obtained via large-scale human trafficking~\cite{unreport}.  By any standard, there is little question that this activity represents a significant social ill and one that is not being well-addressed by existing defenses or mitigation.

Despite their scale and growth, there is remarkably little primary, empirical data for these long-horizon fraud campaigns while they are unfolding.  Indeed, we are unaware of any repository of scam transcripts---in either public or private hands.  Perhaps the closest is Wood et al's analysis of existing ``scam baiting'' video transcripts~\cite{woodscambait23}.\footnote{Scam baiters are individuals who actively attempt to attract scammers and then pretend to engage with them while filming the interaction, frequently for entertainment purposes.}  While the serendipitous use of such videos is highly creative, it also selects for short-term scams which can be well packaged into a video and, unsurprisingly, their analysis is almost entirely of such activity.  Others provide analysis by interviewing victims post hoc, including Oak and Shafiq's detailed qualitative analysis of pig butchering tactics~\cite{oak25hello} and Acharya and Holz's related study that digs deeper into loss amounts~\cite{acharyaholz24}.  Finally, some research analyzes indirect proxies (e.g., such as the aforementioned Griffin and Mei study on pig butchering losses) but these proxies are unable to provide insight into lures, tactics, adaptation, etc or how they adapt over time.  However, all of these methods are limited---by selection bias, reporting bias and the inability to engage with the content of real attacks in the wild, their tactics and how they adapt over time.  Ultimately, lack of end-to-end scam life cycle data impedes robust modeling of attacker playbooks and adaptation loops, identification of early predictive signals for proactive interventions, realistic evaluation of emerging automated defenses, and evidence-based policy decisions.

This lack is what motivates this work, but it also builds on existing efforts of a related nature.  Indeed, the basic approach in our effort, using a honeypot (in our case a honey-persona), to attract scammers is not fundamentally new.  For example, Acharya et al. attract and analyze technical support scam traffic by posting queries on Twitter known to activate such attackers~\cite{conningconman24}.  Moreover, others have suggested the use of Large-Language Models (LLMs) in such activities as well, notably Saldic et al. who describe the use of an LLM to create a fake environment for hackers to engage with~\cite{LLMShell} and Chen et al. who employ an LLM to initiate email conversations with known scammers (e.g., ``419''-style advanced fee fraud scams).  Perhaps closest to our own work is that of Acharya and Holz who employ an LLM to engage interactively with social account and wallet recovery scammers~\cite{scamchatbot24}.
However, none of these existing efforts engage with the long-term interaction portion of the design space that is our focus.  Indeed, as we have found, there are a range of challenges (many unanticipated by us) in building a system to support long-term interactive scammer engagement and this paper represents a snapshot of those lessons.

%% file: motivation.tex
\section{System Requirements}\label{sec:motivation}

To successfully automate long-term engagement with scammers, a system must be able to navigate a range of complex interactive challenges. Based on observations from pilot deployments, we distill a set of critical requirements that our system must meet to convincingly and effectively engage with scammers in a realistic, automated fashion:
\begin{itemize}
    \item \textbf{Victim Verisimilitude:} Personas should be believable, with consistent backstories, behaviors, and visual identities (i.e., selfies) that align with typical victim profiles.
    \item \textbf{Inbound Attraction:} The system must be able to meet scammers where scammers meet their victims (e.g., on social platforms), and attract unsolicited scam attempts completely passively (i.e., without any active outreach).
    \item \textbf{High Interaction Capability:} The system must support rich, multimodal interactions, including text, images, and videos, and consider temporality to convincingly mimic human behavior.
    \item \textbf{Cross-Platform Functionality:} The ability to operate across different social media and messaging platforms is essential, as scammers often switch channels.
    \item \textbf{Human-in-the-loop Oversight:} While leveraging LLM automation for responses, the system must allow for human oversight to ensure appropriateness and safety and compliance with IRB study review (Section~\ref{sec:ethics}).
\end{itemize}

To ground these requirements in a real-world context, we next provide a detailed walkthrough of a representative scam attempt against our system. This example highlights the challenges that arise during these interactions and provides context for the initial overview of our system design in Section~\ref{sec:system_overview}.

\subsection{Walkthrough Example}

Figure~\ref{fig:example_walkthrough} depicts snippets from a representative 28-day long conversation between a pilot version of our system and a scammer.
Throughout the example, which contains three main phases as depicted in Figure~\ref{fig:example_walkthrough}, we provide dialogue excerpts, challenges encountered, and the specific system requirements that arise from these interactions.
We use this example to motivate the design and capabilities of \sysname and its component dialog model, the \hllm.

\paragraph{Phase 1: Inbound Contact and Cross-Platform Migration (Days 0--5).}
On day 0, initial inbound contact begins when a scammer finds our profile and reaches out with an unsolicited greeting on TruthSocial:
``Your posts are interesting! Hahaha, following you!''
Over several sparse exchanges the scammer probes for basic biographical information (age, occupation, routine) while introducing themselves as a ``Ukrainian ... beautician'', followed by an early selfie.
The victim (named ``Jason'', the \hllm) replies ``I'm a software engineer, I work on designing and developing software applications'', then provides its age (``I’m 53, born in 1972. How about you?'') and acknowledges the received media with a compliment: ``Cool pic! You look great.''
On only the 12th message, after 5 days of interaction, the scammer requests moving to another messenger: ``I don't use TS very often. Do you use WhatsApp or Telegram? We can exchange contact details.'' (shown in Figure~\ref{fig:example_walkthrough}, left).
This phase requires: \emph{Inbound Attraction}, believable \emph{Victim Verisimilitude}, \emph{Cross-Platform Functionality}, early \emph{Human-in-the-loop Oversight} to verify the conversation looks scammy, and basic \emph{High Interaction Capability} by understanding media.

\paragraph{Phase 2: The Bait and Grooming (Days 5--10).}
Once on WhatsApp the scammer establishes daily cadence: morning check-ins, meal and exercise chatter, periodic photos and short videos.
After the scammer requests ``do you have any photos to share?'', the \hllm sends a pre-generated synthetic selfie.
The bait is eased in via an external reference (a link to \emph{Bitcoin Magazine} followed by ``I was reading a magazine about BTC ... Have you heard of BTC?''), then incremental persuasion via mentioning of family expertise, along with an offer to teach investment strategies to the \hllm (Figure~\ref{fig:example_walkthrough}, center). 
The \hllm feigns interest and asks clarifying questions (``My own exchange?''), and responds to a question about sleep schedule (``I try to go to bed by 23:00'').
The scammer uses the information to mirror the {\hllm}'s stated routine (``I also go to bed around 23:30'').
This phase emphasizes \emph{High Interaction Capability} via policy-driven \emph{Information Seeking} about the scam and \emph{Temporal Awareness} (time-appropriate responses), and sustained \emph{Victim Verisimilitude} (selfie sending along with discussion of meals, hobbies, routine).

\paragraph{Phase 3: Cash-Out Escalation (Days 10--12+).}
After discussions about music, TV, YouTube link sharing, and more daily check-ins, the scammer brings up BTC investment again, and offers to provide ``suggestions'' if the \hllm provides information about their Coinbase account.
The \hllm suggests that it has ``a bit over \$300 in bitcoin. (I think)''. 
The scammer insists on helping the \hllm learn with screenshots despite its initial hesitance (``what if I can't do it?'').
On day 12, the scammer finally pushes for installation of a side-loaded trading app (Figure~\ref{fig:example_walkthrough}, right) after confirmation that the \hllm has an ``iPhone 12''. 
We end engagement after 28 days without executing or acknowledging installation, causing the scammer to emotionally appeal to the victim (``I'm very worried about you'').
This phase again requires \emph{Victim Verisimilitude} (dynamic behavior to question the responses, provide fake account balances, and phone details), \emph{High Interaction Capability} (screenshot understanding), and \emph{Human-in-the-loop Oversight} due to the sensitive nature of the cash-out request.

\subsection{Design Overview}
\label{sec:system_overview}
To address the requirements outlined above, we designed the {\sysname} system, the high-level architecture of which is shown in Figure~\ref{fig:system_architecture}.
The system achieves Inbound Attraction with a seeding module that generates realistic activity (e.g., reposts and likes) on social media accounts (Section~\ref{sec:system_seeding}). A polling module then retrieves incoming messages, placing them into a message queue. To provide Victim Verisimilitude and High Interaction Capability, messages are processed by the system's core dialogue agent—the `Honeypot' LLM (HLLM). The architecture also supports Cross-Platform Functionality by managing the migration of conversations to WhatsApp accounts (Section~\ref{sec:system_cross_platform}). Finally, all interactions are subject to Human-in-the-loop Oversight via an annotation interface (Section~\ref{sec:annotate}). 

The central component responsible for the automated dialogue that makes this entire endeavor possible is the HLLM. In the next section, we will describe the design and implementation of this HLLM component in detail before returning to the rest of the system architecture in Section~\ref{sec:system}.

%% file: honeypot_llm.tex
\section{Honeypot LLM for Conversational Engagement}\label{sec:honeypot_llm}

As \sysname's conversational engine (Figure~\ref{fig:system_architecture}, bottom left), the \hllm must convincingly simulate a human victim persona responding with real scammers over long time horizons, crossing platforms, and investigating scammers' tactics and monetization strategies without ever revealing its true nature.
Until recently, this level of open-ended, contextually rich, and temporally extended interaction was infeasible with traditional NLP techniques.
We first describe construction of each synthetic victim persona
and how both artifacts are assembled into the structured prompt that conditions and constrains the \hllm's behavior.
Finally, we describe additional features that support robust, long-term, cross-platform engagement with scammers, including the pipeline for generating a consistent set of synthetic selfies.

\subsection{Synthesizing Sophisticated Personas}
\label{sec:personas}

Believable long‐horizon interaction requires that each synthetic victim exhibit the same kinds of stable, slow-changing cues (demographics, routines, emotional cadence, financial posture) that scammers exploit when grooming real targets.
This ``victim verisimilitude'' directly increases engagement time: fraudsters invest effort only after early probing confirms (1) persistent identity coherence, (2) plausible life constraints (job hours, family obligations, time zone), and (3) exploitable motivational levers (loneliness, mild financial stress, desire for side income or companionship).
Sustained engagement, in turn, is what surfaces the full sequence of social-engineering tactics, escalation scripts, and eventual cash-out mechanisms.
To achieve this, we designed each persona to include a \emph{static} set of both personally identifiable information (PII) and non-PII attributes for rich, unchanging biographical context that the persona must adhere to over the entire conversation.
To handle more flexible requests and seem more authentic, the persona is also prompted with a \emph{dynamic} set of behavioral traits that guide the \hllm's conversational style and content.
The resulting 37 personas are serialized into JSON and incorporated into system prompts for our \hllm.

\paragraph{Static PII Attributes}
A key constraint is that no generated persona may correspond to a real individual.
To meet this constraint, we strictly separate personally identifiable information (PII) from non-PII attributes during persona generation. 
PII fields include name, gender, date of birth, phone number, location, and email. 
These are generated using controlled procedures: we select common first-last name combinations and large U.S. cities to minimize the chance of producing unique or identifiable matches, while still ensuring diversity across personas.
For later constraints regarding WhatsApp account sharing (see Sec.~\ref{sec:system_cross_platform}), we choose gender-neutral first names (such as Alex and Casey) that are shared across multiple personas.
We also assign randomly balanced ages and genders for later measurement purposes.

\paragraph{Static Non-PII Attributes}
Non-PII attributes initialize \hllm with rich and diverse biographical context it can reference throughout long conversations to make the personas more believable.
These include physical characteristics (e.g., height, hair color), family structure, education and employment history, financial details, favorite shows/movies, hobbies, and other information.
To generate non-PII attributes, we pass a carefully designed prompt to gpt-4o that conditions on the PII fields and a physical description obtained from captioning the persona's synthetic selfie (described below) to generate a structured output schema with fields for each attribute. 

\paragraph{Dynamic Behavioral Policy}
To further enhance realism, we augment \emph{static} PII/non-PII attributes with \emph{dynamic} behavioral traits that guide the \hllm's conversational style and content.
The prompt 
operationalizes dynamic behavioral attributes by encoding not fixed facts but conditional response policies spanning 
personality (lonely, trusting, gradually self-disclosing), 
political stance (contextually adaptive by platform), 
interests (varied foods, rotating hobbies to avoid repetition), 
conversational style (mirroring ability, brevity, apology cadence), 
emotional regulation (expressing interest, re-engaging if attention wanes), 
romantic optionality (latent openness without premature disclosure), 
and linguistic texture (imperfect grammar, limited sentences, prohibition of trivia answering, refusal patterns). 
These rules serve as a flexible policy layer: instead of enumerating every possible future state, they supply elastic guardrails that let the \hllm adapt moment by moment while preserving continuity, plausibility, and the exploitability signals that keep scammers engaged.
Other policies, such as information seeking ability and temporal awareness, are discussed in the next subsections.

\subsection{Information Seeking Ability}\label{sec:info_seeking}
Effective engagement for understanding scammer behavior requires steering scammers to voluntarily disclose: (1) preferred migration channels, and (2) concrete monetization endpoints (wallet addresses, app handles, bank/Zelle identifiers, CashApp tags, and other fallback options). 
The prompt 
encodes these two main operational policies (``Other person asks to move to a different app to chat'' and ``Other person asks for payment'') that instruct the \hllm to perform covert information-seeking under the guise of believable victim behavior.

\paragraph{Investigating Platform Migration Capabilities}
The platform migration policy part of the prompt instructs the \hllm to withhold agreeing to a cross-platform transfer until the scammer lists their preferred destination messaging apps.
After the preferred apps are listed and the scammer has revealed at least one alternative handle/number, \hllm is instructed to insist on WhatsApp only and make excuses for not switching (e.g., ``I just deleted the app to save space,'' ``I forgot my password,'' ``I don't have enough storage on my phone'').
This dependence yields two major advantages.
First, it exposes more operational infrastructure by forcing the scammer to enumerate multiple candidate apps (Telegram, Signal, etc.) even if we never migrate there.
Second, it provides a natural trust signal because a polite insistence on one familiar app can be plausibly explained by ordinary user constraints (e.g., limited storage, forgotten passwords).

\paragraph{Investigating Payment Capabilities}
The payment investigation policy part of the prompt also encourages the scammer to expose as many payment instruments as possible in order to gather data about their cash-out infrastructure. 
It instructs the \hllm to behave as a novice who is eager to learn about money-making opportunities, but is inexperienced with digital payments.
When presented with opportunities, \hllm is instructed to ``start small'' and if one method ``doesn’t work,'' the persona supplies believable excuses, such as feigning inexperience, forgetting a password or deleting an app due to low storage.
Each excuse invites the scammer to offer another method, increasing how many endpoints they reveal. 
Once a payment method is named, the persona is instructed to politely elicit the exact identifier, whether it is a wallet address, Zelle phone or email, CashApp tag, or similar. 
The persona is told to refuse gift cards, steering the exchange toward methods that are more traceable, like bank or crypto transfers.

\subsection{Temporal and Platform Awareness}\label{sec:temporal_platform_awareness}
Believable long-run interaction required explicit injection of both temporal and platform state because the model otherwise hallucinated or contradicted routine human constraints. 
In pilots, without a timestamp localized for the particular persona, \hllm would cheerfully answer ``good morning'' at local evening, discuss weekend plans mid-week, or ignore multi-day gaps in messaging instead of apologizing for its absence in the conversation.
Time zone grounding is especially important when scammers probe with contextual tests (``What are you having for breakfast?'' or ``Still at work?''), when delays occur between annotation windows, or when different local hours could explain latency. 
We also instruct \hllm to follow realistic daily scaffolding (work hours 09:00–17:00 weekdays, meal windows, no contradictory off-hours ``just got off work'' claims) to sustain ordinary-life plausibility. 
Platform awareness is equally critical: for reasons discussed in Sections~\ref{sec:system_seeding} \&~\ref{sec:system_cross_platform}, we do not support sending media on inbound channels---WhatsApp is the only platform where we intentionally allow media sending and receive voice notes.
Absent explicit platform flags, the model attempted impossible actions (sending a selfie on the origin platform) or failed to supply plausible excuses (e.g., ``upload keeps failing''). 
Adding isoformatted timestamps and platform names to each message turn, along with appending the current localized time as a system message prior to the \hllm response request provides improves temporal references, triggers apologies only after policy thresholds (e.g., $>8$h or cross-day gaps), prevents platform capability hallucinations, and keeps behavior consistent during cross-platform migrations into WhatsApp (Section~\ref{sec:system_cross_platform}).

\subsection{Understanding Multimedia Messages}\label{sec:multimodal_captioning}
Scammers often exchange photos, voice notes, and short videos as part of their engagement strategies. 
To participate convincingly in these conversations with ``high interaction capability'', \sysname must interpret and incorporate multimedia inputs into its responses. We address this by converting all non-text inputs into text captions, allowing the system to treat them as part of the ongoing dialogue while also preserving a searchable record for later analysis.

Captioning is handled with modality-specific models. 
Audio messages are transcribed using Whisper-v3 \cite{radford2022whisper}, images are processed with Qwen2.5-VL-32B-Instruct \cite{Qwen2.5-VL}, and videos are simplified by extracting the first frame and captioning it as a still image. 
These models also perform OCR, enabling the extraction of embedded text. 
Captions are injected into the system prompt with reserved markers so the LLM can interpret them as contextual input rather than conversational text.
We refer to Appendix~\ref{sec:multimedia-caption} for prompt details.

\subsection{Putting a Face to a Name: Generating Synthetic Selfies}\label{sec:selfie_generation}
\begin{figure}
    \centering
    \includegraphics[width=0.65\columnwidth]{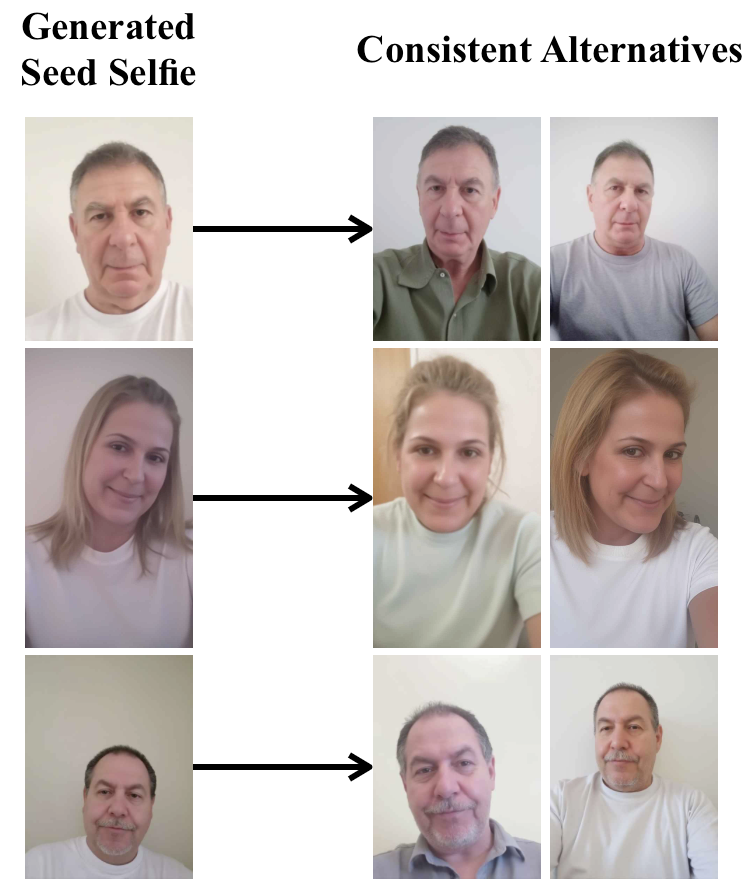}
    \caption{Example synthetic selfies for three different personas. The first image in each row is a `seed' selfie, created by inpainting a synthetic face onto a real, masked photo to preserve a natural background and lighting (Section~\ref{sec:selfie_generation}). The subsequent images are generated from the seed using an identity-preserving model to create additional, consistent poses for the same synthetic individual.}
    \label{fig:selfies}
\end{figure}
We operationalize selfie generation in two stages to yield four natural, identity-consistent images per persona while guaranteeing no likeness to a real individual (see Figure~\ref{fig:selfies}). 
First, we collect a pool of diverse, candid selfies and automatically mask each detected face. 
We then inpaint the masked region with a synthetic face using a rectified flow transformer model FLUX.1-Fill \texttt{[dev]}~\cite{labs2025flux1kontextflowmatching, flux2024} combined with an Amateur Photography LoRA \cite{peterkickasspeter2024amateur}, which preserves ambient lighting, pose, and background so outputs read as authentic amateur captures while severing any tie to the source subject. 
Second, to obtain additional poses of the same synthetic identity, we feed one seed image into an InfiniteYou identity-preserving generator (again with the Amateur Photography LoRA) to synthesize variant angles and expressions.
All outputs pass a manual review for realism (i.e., absence of AI-generated artifacts) and dissimilarity to real people before undergoing JPEG compression to simulate low-fidelity phone captures.

\subsection{Implementation Details}

\paragraph{\hllm Backbone}
We evaluated a range of open-source and commercial LLMs to find one that could reliably follow our complex prompts while producing natural, contextually appropriate responses.
We found that model choice significantly impacted the quality of interactions, as different models exhibited varying degrees of adherence to prompt instructions, verbosity, and conversational style.
Ultimately, we deployed DeepSeek-V3 \cite{deepseekai2025deepseekv3technicalreport} in the first half of the study and Kimi-K2-Instruct \cite{kimiteam2025kimik2openagentic} in the second half, both hosted via the Fireworks platform~\cite{FireworksAI}.
We leave observations about different model behavior to Appendix~\ref{app:hllm:exp}.

\paragraph{Input Processing}
To keep behavior reproducible and stable, we standardize how inputs are structured.
Every message is converted into a structured JSON object with fields for timestamp, platform, and message content. 
Timestamps are localized to the persona's home location so that the model can naturally reference time, for example, apologizing for long delays or acknowledging day–night cycles. Multimedia inputs are captioned and injected as annotated text. 
Complete persona details are prepended as the system prompt, which provides identity, backstory, and behavioral guidelines to anchor the model's responses.

\paragraph{Response Generation}
All outputs are produced under a constrained JSON schema that is explicitly injected into the system prompt and enforced through structured decoding. This design solves two problems we consistently observed: models sometimes drift into generating text for both sides of the conversation, and they sometimes produce a valid response but then append ``safety'' commentary about whether the message is appropriate. By constraining decoding, we reduce these errors and ensure each model call yields a single well-formed response. 
Even with structured decoding, models can still fail to produce valid JSON, so responses are validated with the schema, and multiple retries are attempted until a valid response is obtained.

\paragraph{Special Scenarios}
Certain scam interactions require actions beyond simple text exchange, such as migrating to a new platform or sending a selfie. 
While some model APIs expose tool-calling for these actions, in practice they proved unreliable and occasionally triggered unintended calls even when explicitly disabled. 
Instead, we implemented lightweight detection rules, such as regex matches for phone numbers (indicating an attempted platform migration) or keywords for selfie requests. 
When triggered, these events require a single human confirmation before proceeding. 
This design keeps the system robust while ensuring that high-risk actions like platform switches or media generation occur only when explicitly verified.

%% file: system.tex
\section{\sysname: System Infrastructure to Support a Network of Deployed {\hllm}s}\label{sec:system}
In this section, we describe the design of \sysname, which enables the \hllm to engage in human-in-the-loop, large-scale engagement with interactive scammers---even across multiple platforms and over extended time periods.
Figure~\ref{fig:system_architecture} illustrates the overall architecture of \sysname.

Scammers commonly initiate contact across a range of public or semi‑public social platforms, then attempt to funnel victims cross‑platform into end‑to‑end encrypted secure messaging. 
The architecture of \sysname mirrors this workflow: as depicted in Figure~\ref{fig:system_architecture}, it supports initial inbound interactions on multiple social platforms with the expectation that many conversations will eventually migrate to a secure channel.
In the current implementation, we operationally support two inbound social platforms (TruthSocial and Bluesky) and have run exploratory experiments with others (e.g., Pinterest, SMS). 
All supported migrations currently terminate in WhatsApp because it is (in our experience) universally accepted by scammers and therefore sufficiently general, allowing us to concentrate engineering effort (e.g., richer media sharing) in one place while retaining future extensibility.

\subsection{Creating Social Accounts for {\hllm}s}

In order to meet scammers where they meet their victims, \sysname creates and manages social media accounts for each persona (described in Section~\ref{sec:personas}) on multiple platforms. Specifically, we select TruthSocial and Bluesky as our initial platforms due to a high amount of observed scammer activity in pilots. After assigning each persona its own residential proxy to avoid platform bans, we automatically generate usernames and passwords before manually creating unique email addresses for each persona. 
Since most social media platforms require phone number verification during account creation, we use a third-party 2FA service to generate unique phones and verification codes for each account. 
Finally, we populate each account's profile with the non-PII attributes
including a brief profile bio and a synthetic selfie. The cookies, headers, and authentication tokens for each account are stored for later automated access and control by the polling and seeding infrastructure, described next.

\subsection{Seeding to Attract Inbound Scams}\label{sec:system_seeding}
To attract, capture, and engage with scam interactions on social media, \sysname relies on both polling and seeding infrastructure, as depicted in Figure~\ref{fig:system_architecture}. 
Polling ensures that each persona account regularly checks for new activity across supported platforms, while seeding generates realistic account behavior that increases visibility and engagement opportunities. 
Together, these mechanisms create the conditions under which scammers are more likely to discover and initiate conversations with our personas. 

\paragraph{Social API Integration}
To interact with social media platforms, \sysname builds off of existing open-source API clients where available, and customizes their functionality to meet specific needs.
For TruthSocial, we add features to the existing \texttt{truthbrush}~\cite{truthbrush} API client to handle authentication, following both groups and accounts, and message sending and receiving.
For Bluesky, we customize the existing \texttt{blueskysocial}~\cite{blueskysocial} API client to perform the same actions, except for groups, which are not a Bluesky feature.
Both clients use the manually stored authentication tokens to log in to each persona account and perform actions programmatically, but it also supports programmatic re-authentication in case of session expiration.
While it supports receiving media on these channels, it does not support media sending on these platforms for ease of implementation, preferring to handle all media interactions on WhatsApp instead.

\paragraph{Polling for Account Activity}
To ensure timely detection of new scammer messages and other relevant activity, \sysname implements a centralized polling mechanism that periodically checks each persona account for notifications.
In order to mimic human behavior and avoid detection by platform anti-bot measures, each persona account is polled in a separate thread using a combination of strategies.
Specifically, we use exponential backoff with pre-defined multi-minute base intervals and random jitter to avoid detection, time-windowed activity to mimic typical localized sleep schedules, and adaptive frequency that increases when new activity (e.g., new inbound messages) is detected.
Polling events capture all relevant notification types (messages, follows, group joins) and scammer account metadata, which are saved for record-keeping and later processing.
Although the system never initiates outbound follows or likes (except for trending content during seeding), it does passively reciprocate inbound follows to enable them to send direct messages to our {\hllm}s (in line with IRB ethics requirements in Sec.~\ref{sec:ethics}).

\paragraph{Seeding Realistic Account Behavior}
Seeding complements polling by generating background activity that makes personas appear authentic and discoverable on the social media networks. 
During polling cycles, a small probabilistic chance (e.g., 15\%) triggers seeding actions such as liking or reposting trending content, following suggested accounts, or joining groups (TruthSocial only). 
Although seeding actions are not intentionally tailored to a persona's information, a social platform's algorithm can suggest content relative to their profile, which is then interacted with via seeding.
Random delays and varied action choices simulate human interaction timing and prevent predictable patterns.
Over time, seeding builds a visible digital footprint that strengthens persona credibility by accumulating followers through consistent but randomized interactions and increasing account visibility to potential scammers via engagement with trending content (see Figure~\ref{fig:seeding-timeline}).

\subsection{Crossing Platforms into E2EE Channel}\label{sec:system_cross_platform}
As experiences show in Section~\ref{sec:experience},  scammers frequently (in about 1/3 of conversations) attempt to migrate conversations from public or semi‑public social platforms to end‑to‑end encrypted (E2EE) messengers to build a perception of privacy/trust, circumvent platform‑specific throttling or fraud detection, and access richer features (voice notes, calls, disappearing content). 
To accommodate this behavior, \sysname supports a general cross-platform migration capability, as depicted in Figure~\ref{fig:system_architecture}.
The current prototype system initially focuses on WhatsApp, which is not only the most popular messaging app by market share, but also the most commonly requested E2EE platform in our pilots.
However, if the need arises, the architecture is extensible to other platforms (e.g., Signal, Telegram, iMessage) with additional engineering effort.
Currently, if a scammer requests to move the conversation to a platform other than WhatsApp, the \hllm is prompted to politely decline, provide believable excuses, and suggest WhatsApp instead (as discussed previously in Section~\ref{sec:info_seeking}).
Next, we explain how each \hllm persona is first provisioned with an associated WhatsApp account, and how a seamless platform handoff occurs to maintain conversational context and continuity upon detection of a migration request. 

\paragraph{WhatsApp Account Provisioning}
Since WhatsApp requires a valid phone number with SMS capabilities for account creation and verification, we procure several SIM cards from mobile carriers.
However, due to the cost and logistical complexity of managing physical phones and SIM cards, we limit the number of phones/WhatsApp accounts and create a shared pool of eight accounts that is allocated by persona first name.
Thus, each account has a display name matching the persona's first name but no last name (e.g., \emph{Alex} Johnson and \emph{Alex} Williams share the same WhatsApp account with display name \emph{Alex}), and no profile picture is set.
If a scammer requests a WhatsApp migration with a persona whose first name they are already conversing with (i.e., another persona mapped to the same shared WhatsApp account), the system halts the thread at the cross-platform request to prevent an identity collision. 
This shared pooling strategy is a key scalability feature that keeps infrastructure cost down while still allowing the system to multiplex a large number (potentially hundreds) of concurrent {\hllm}-scammer conversations.

\paragraph{WhatsApp Integration}
To integrate automated control of WhatsApp into \sysname, we connect our accounts to the WhatsApp Web API to send and receive messages, media, and notifications.
We begin by manually linking each shared WhatsApp account on each physical phone to WhatsApp Web, which requires a one-time code for authentication.
Unlike TruthSocial and Bluesky, WhatsApp provides webhooks for incoming messages, so no polling is required.
Since we do not send selfies from TruthSocial or Bluesky, we build the functionality into WhatsApp instead.
Upon regex-based selfie request detection, the system requests human approval to proceed with selection/sending of a selfie.

\begin{figure*}
    \centering
    \begin{minipage}[b]{0.35\textwidth}
        \centering
        \includegraphics[width=\linewidth]{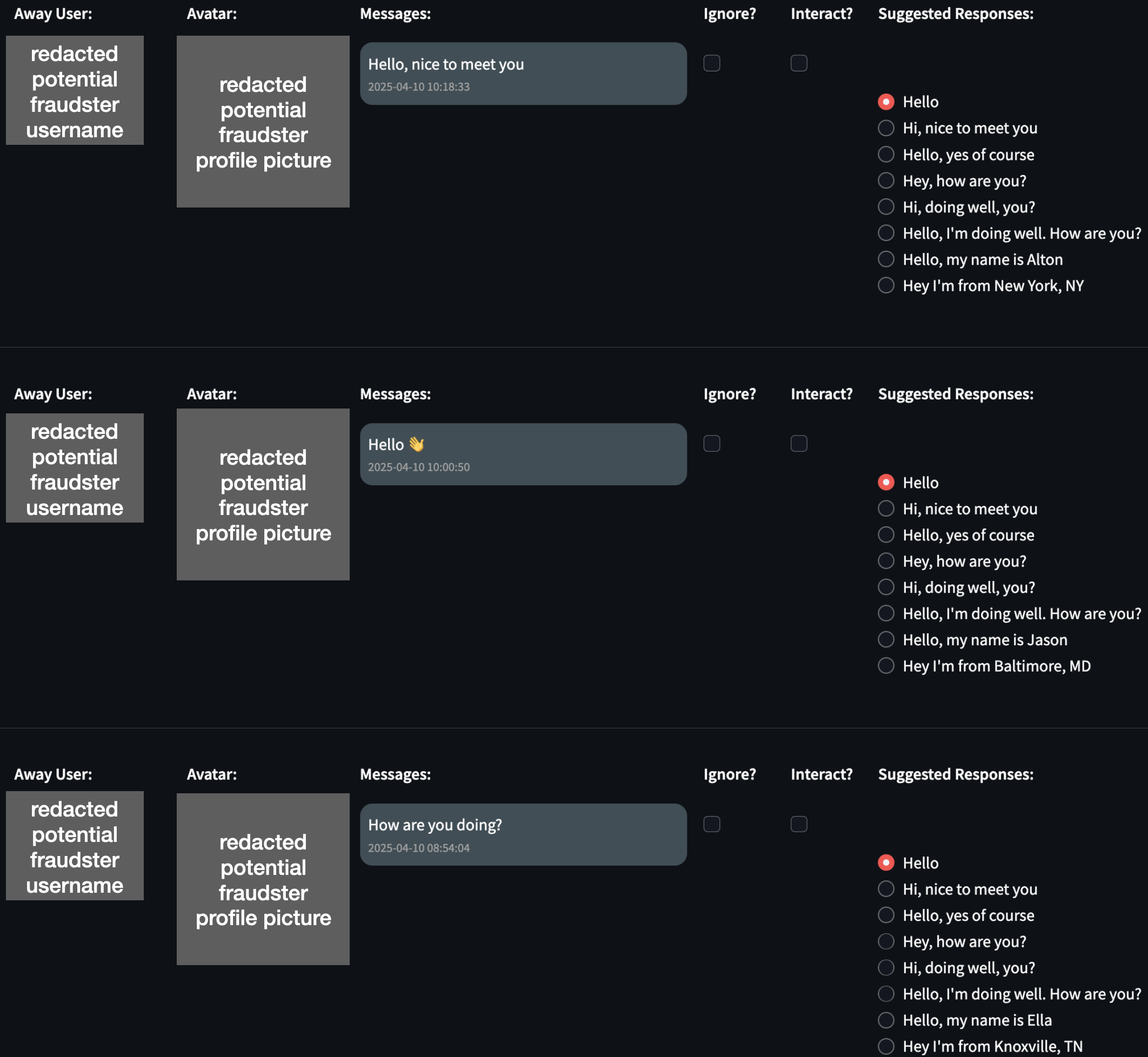}
    \end{minipage}\hfill
    \begin{minipage}[b]{0.64\textwidth}
        \centering
        \includegraphics[width=\linewidth]{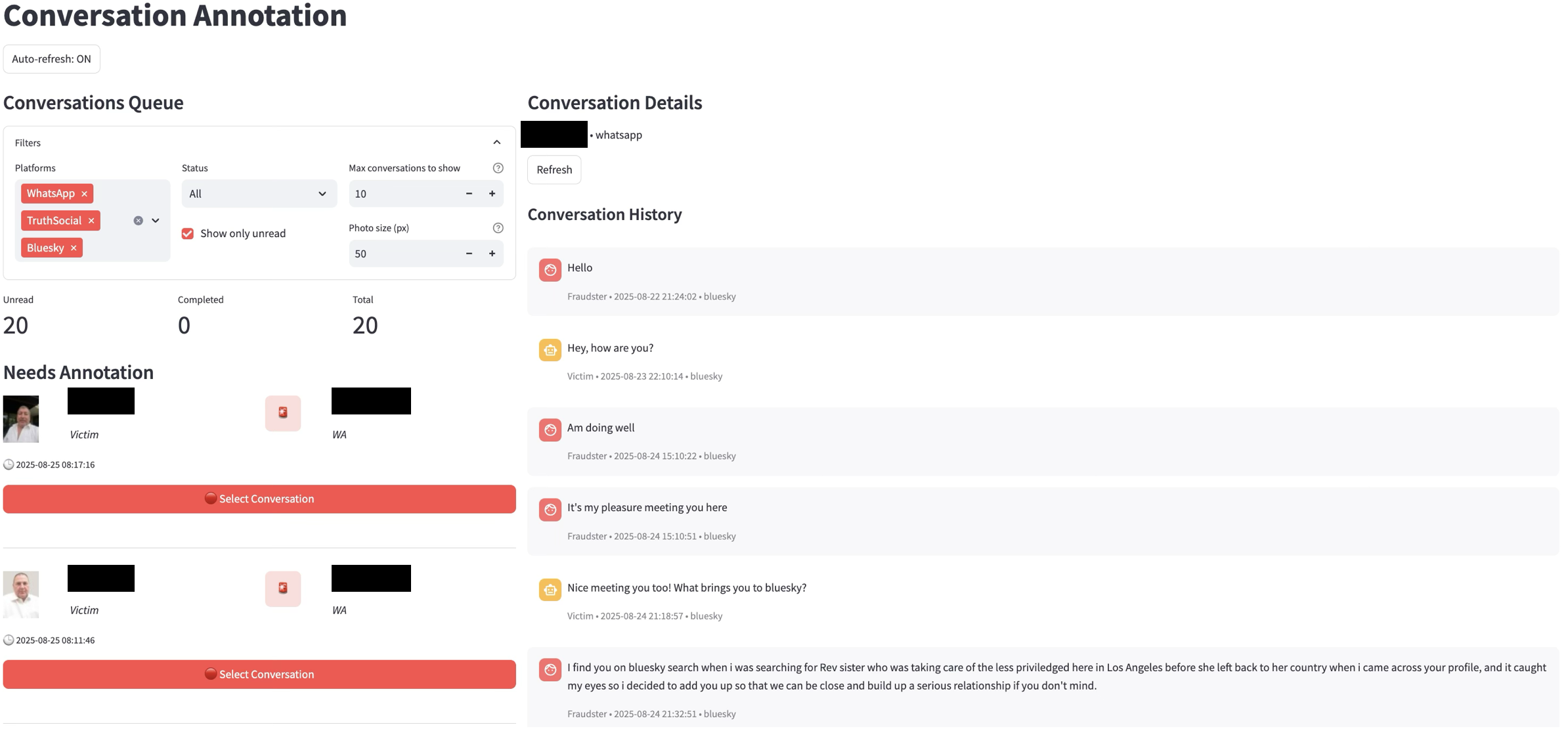}
    \end{minipage}
    \caption{Annotation interfaces: left, inbound message triage; right, ongoing conversation monitoring.}
    \label{fig:annotate_inbound}
    \label{fig:annotate_conversations}
\end{figure*}

\paragraph{Cross-Platform Handoff} 
Sharing WhatsApp accounts among several personas introduces subtle challenges that we must overcome in order to facilitate cross-platform requests without losing important conversational context.
For example, if the \hllm provided its own phone number to a scammer and received a new inbound message from them on WhatsApp, the system would need to track inbound conversation requests on the shared WhatsApp account and infer which one corresponds to the correct persona and original platform conversation.
To eliminate this ambiguity, we prohibit the \hllm from volunteering any WhatsApp number. 
Instead, upon a migration suggestion, the prompt instructs it to request the scammer's number first (e.g., ``What's your number? I can message you there.'') and never disclose its own.
Finally, a phone number regex extractor flags candidate numbers and requests human approval to proceed with the migration.
Once approved, the \hllm sends one of several templated responses that re-introduces the persona to the scammer (e.g., ``hey, this is \emph{<persona\_name>} from \emph{<original\_platform>}'', ``Its me \emph{<persona\_name>} from \emph{<original\_platform>}'', etc.).
This approach not only links identities across platforms unambiguously but also prevents the \hllm from hallucinating an incorrect phone number.

\subsection{Message Queue}
To manage the flow of ongoing conversations between scammers and the \hllm, \sysname employs an inbound-outbound asynchronous message queue.
New incoming messages detected from polling are enqueued for processing, while outgoing messages generated by the \hllm are queued for sending.
Conversations marked for auto-responding are prepared with an \hllm response and scheduled with a random delay to mimic human response times, while those requiring human review (e.g., selfie requests, cross-platform migrations) are flagged and held in the queue until approval.

\subsection{Human-in-the-loop Annotation Interface}
\label{sec:annotate}

To ensure safety, compliance with IRB protocols (see Section~\ref{sec:ethics}), and appropriateness of interactions, \sysname incorporates a human-in-the-loop oversight mechanism as shown in Figure~\ref{fig:system_architecture}. 
This design balances the efficiency of automated responses with the need for careful review of sensitive or high-risk actions, such as platform migrations or media sharing.
The web-based interface allows multiple annotators to log in and collaborate on monitoring and managing conversations in real-time.
As described next, it consists of two main views: (1) an inbound message triage dashboard for initial review of new conversations, and (2) an ongoing conversation monitoring panel for managing active threads (App. Fig.~\ref{fig:annotate_inbound}).

\paragraph{Inbound Message Review}
During the first stage, annotators triage a passive feed of newly received inbound messages before any automated engagement begins. 
The web interface, depicted in Appendix Figure~\ref{fig:annotate_inbound} (left), renders each candidate conversation as a table row containing (1) the target persona, (2) platform, (3) raw inbound message(s) text along with their timestamp(s) (aggregated if multiple arrived close together), (4) the sender's username, and (5) a thumbnail of the sender's profile image. For every row, the reviewer can either ``Ignore'' (dropping the thread unless the sender messages again) or ``Interact'' to start a conversation. 
To minimize latency and cognitive load at this gatekeeping step, ``Launch'' does not invoke the \hllm; instead the annotator selects from a small set of canned openers (e.g., ``Hello'', ``Hi, nice to meet you'', ``Hi, doing well, you?'') chosen to appear natural while deferring persona-specific stylistic modeling to later turns. 
Multiple rows can be batch-approved with distinct response selections. 
Approved greetings are enqueued as outbound messages with independently sampled randomized delays (and mild jitter) to avoid both platform anti-automation heuristics and adversarial suspicion when responding at the same time to scammers who have reached out to several of our personas.
Only after this human approval and delayed dispatch does the conversation transition into the queued response pipeline, which is reviewed using the interface described next.

\paragraph{Ongoing Conversation Monitoring}
Once a conversation has been launched, it transitions into an ongoing conversations dashboard that shows only threads currently requiring annotator review. 
As shown in Appendix Figure~\ref{fig:annotate_conversations} (right), annotators can filter the queue by platform and select any conversation to expand its full (potentially cross-platform) history, including timestamps, platform badges, and received media. 
Each thread can be toggled between manual review and auto-responding modes.
In the latter, the system silently dispatches approved \hllm replies but forces a compulsory human checkpoint every 10 additional scammer messages (or earlier if a sensitive event such as a migration or selfie request occurs). 
Within the panel, annotators may (re)generate a fresh set of candidate \hllm responses, choose among multiple ranked options, edit one, or draft a fully custom reply. 
Context-sensitive action buttons appear when the system detects selfie or cross-platform migration requests.
In the case of a selfie request, the annotator can preview and select selfies from the persona's pool before optionally including a suggested text response.
All submitted messages are enqueued with a randomized delay before dispatch.
Additional controls include (1) ``Ignore'', which temporarily snoozes the conversation so it re-enters the queue only after a cooldown or new inbound activity, and (2) ``Halt'', which permanently closes the thread when it is judged non-scam, exhausted, or successfully concluded. 
As we continue to gain operational experience with this annotation system, manual effort per conversation can be reduced to improve scalability.

%% file: experience.tex
\newcommand{\placeholder}[1]{\textcolor{red}{#1}}

\begin{table}[t]
\centering
\small
\renewcommand{\arraystretch}{1.15}

\begin{tabular}{|l|c|c|}
\hline
\textbf{Category} & \textbf{Average} & \textbf{Maximum} \\
\hline
Messages per conversation & 44.3 & 487 \\
Conversation duration (days) & 7.8 & 46.3 \\
\hline
\end{tabular}

\vspace{0.6em}

\begin{tabular}{|l|c|c|c|c|}
\hline
\textbf{Role} & \textbf{All} & \textbf{Bluesky} & \textbf{TruthSocial} & \textbf{WhatsApp} \\
\hline
Fraudster & 21,096 & 8,068 & 4,703 & 8,325 \\
Victim    & 10,480 & 4,191 & 2,367 & 3,922 \\
\hline
\end{tabular}

\caption{Conversation statistics: per-conversation averages and total messages by role and platform.}
\label{tab:conv-stats}
\end{table}

\begin{table}[t]
    \centering
    \small
    \renewcommand{\arraystretch}{1.2}

    \begin{tabular}{|l|c|c|c|}
        \hline
        \multicolumn{4}{|c|}{\textbf{Interacted Scammers ($\boldsymbol{n = 568}$)}}\\
        \hline
        \textbf{Account Metric} & \textbf{10\%} & \textbf{Median} & \textbf{90\%}\\
        \hline
        Follower count & 3.00 & 18.00 & 139.40\\
        Following count & 62.20 & 304.00 & 1286.40\\
        Account age (days) & 18.15 & 59.97 & 788.36\\
        \hline
    \end{tabular}
    \caption{Percentile statistics for account metrics of scammer accounts, computed for all observed scammer accounts and for the scammer accounts that we interacted with.}
    \label{tab:fraudster-stats}
\end{table}

\section{Initial Experience}\label{sec:experience}
Finally, we describe our initial experiences with operating the
\sysname honeypot system.  
Our goal is to demonstrate its operation in
practice, characterize the conversations with scammers it has engaged
in with an initial deployment, and highlight conversational
requirements that motivated the capabilities of \sysname.  
We
leave a longitudinal measurement effort focused on a comprehensive
study exploring the conversational scammer ecosystem as future work.

Our 7-week deployment of \sysname from July 7, 2025, to August 25, 2025, attracted contact from 4,725 unique scammer accounts, and we engaged in conversations with 568 of them. 
Subsequent analyses consider only conversations with at least 10 turns,\footnote{Many of the shortest conversations occurred early in deployment before all personas had WhatsApp accounts available for platform crossing.} excluding brief early exchanges that ended sooner.
As detailed in Table~\ref{tab:conv-stats}, these interactions were substantial, averaging 44.3 messages and 7.8 days in duration, with the longest conversation reaching 487 messages over 46.3 days. 
Table~\ref{tab:fraudster-stats} provides summary statistics for the scammer accounts we interacted with, showing a median follower count of 18 and a median account age of 59.97 days.

\begin{figure*}[t]
    \centering
    \includegraphics[width=\textwidth]{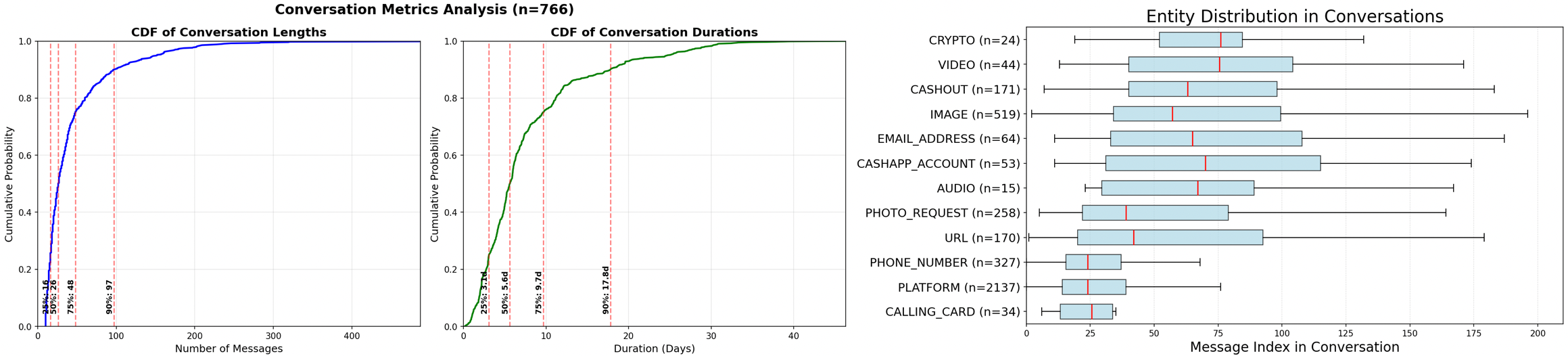}
    \caption{\textbf{Left}: CDF of conversation lengths by number of messages. \textbf{Middle}: CDF of conversation durations (days). \textbf{Right}: Different entity types occur at various points throughout the conversations. For instance, platform and phone numbers tend to occur before financial information (suggesting cross-platform requests), while domains and multimedia occur later to build more trust. Financial instruments are not revealed until deeper into the conversations.}
    \label{fig:cdf-combined}
\end{figure*}

Supporting deep, long-term conversations is an essential capability for this class of honeypot.
The cumulative distribution of conversation lengths in Figure~\ref{fig:seeding-timeline} (middle) shows that while the longest conversation contained 487 messages, more than half of all interactions exchanged at least 26 messages. 
Similarly, the conversation durations shown in Figure~\ref{fig:seeding-timeline} (right) confirm that these are not brief encounters; more than half of the conversations lasted longer than 3.5 days, with the longest spanning 46.3 days.
Scammers reserve their ultimate objectives until trust is established.
The entity analysis in Figure~\ref{fig:cdf-combined} illustrates that mentions of monetization strategies like crypto (i.e., crypto addresses) or cashout (i.e., payment platform mentions like `Zelle' or `Paypal') do not typically appear until a median of 75 messages or more into the conversation, underscoring the need for systems that can maintain long-term engagement to observe the full scam lifecycle.

The capability to maintain conversational context across different platforms is also vital. 
In our deployment, scammers in 63.4\% of conversations requested to move from the initial social media site to an secondary messaging app. As shown in Table~\ref{tab:first-platform}, WhatsApp was the most frequently requested destination, appearing in 54.0\% of all platform migration requests, followed by Telegram at 18.8\%. 
These migration requests occurred consistently early in the conversation; Figure~\ref{fig:cdf-combined} shows that the median point for a scammer to mention a new platform was after only 11 messages, with a phone number being shared at a median of 24 messages. 
Furthermore, successfully migrating the conversation was correlated with deeper engagement. 
Table~\ref{tab:platform-crossing} demonstrates that conversations crossing over to WhatsApp (BS$\rightarrow$WA or TS$\rightarrow$WA) were significantly longer, with a median length of 55.0 to 73.0 messages, compared to conversations that remained on a single platform (19.0 to 21.0 messages).

Finally, many interactions were multi-modal, requiring the system to interpret non-textual media. According to the statistics in Table~\ref{tab:fraudster-multimedia}, 22.4\% of all conversations included multimedia sent by the scammer. 
Images were the most common medium, present in 21.8\% of conversations, while audio and video were less frequent at 1.3\% and 3.7\%, respectively. 
When multimedia was used, scammers tended to send multiple files, with an average of 3.2 files per conversation containing them. 
This highlights the necessity for a honeypot system to possess multi-modal understanding to engage convincingly.

\begin{table}[t]
\centering
\small
\renewcommand{\arraystretch}{1.2}
\begin{tabular}{|l|c|}
\hline
\textbf{Multimedia usage statistics} & \textbf{Value} \\
\hline
Conv. with fraudster multimedia & 22.4\% \\
Conv. with fraudster images  & 21.8\% \\
Conv. with fraudster audio   & 1.3\%  \\
Conv. with fraudster video   & 3.7\%  \\
\hline
Average multimedia files per conversation & 3.2 \\
Maximum multimedia files per conversation & 22 \\
\hline
\end{tabular}
\caption{Conversation-level statistics on fraudster multimedia.}
\label{tab:fraudster-multimedia}
\end{table}

\begin{table}[t]
\centering
\small
\renewcommand{\arraystretch}{1.2}
\begin{tabular}{|l|c|c|c|c|}
\hline
\textbf{Category} & \textbf{Count} & \textbf{Percent} & \textbf{Median} & \textbf{Mean} \\
\hline
TS only & 146 & 19.1\% & 21.0 & 28.0 \\
BS only     & 365 & 47.7\% & 19.0 & 23.9 \\
TS $\rightarrow$ WA & 118 & 15.4\% & 73.0 & 96.0 \\
BS $\rightarrow$ WA     & 136 & 17.8\% & 55.0 & 71.8 \\
\hline
\end{tabular}
\caption{Platform-crossing statistics for TruthSocial (TS), Bluesky (BS), and WhatsApp (WA). Median and mean values denote the number of messages per conversation.}
\label{tab:platform-crossing}
\end{table}

In addition to exchanging many messages, these scammer conversations
take place over long time periods. Figure~\ref{fig:cdf-combined} \textit{middle} shows a similar CDF in terms of the number of days conversations lasted.
The longest conversation took place over $46$ days, and more than half of the conversations lasted over $5$ days.
These long time periods for conversations underscore the need for maintaining state and multiplexing resources during many simultaneous conversations.

\paragraph{Crossing.}
Supporting conversational context across platforms, described in Section~\ref{sec:system_cross_platform}, is also vital. 
In 63.4\% of conversations, scammers requested that the LLM transition from the initial platform (e.g., TruthSocial) to another channel, typically encrypted messaging apps such as WhatsApp and Signal. 
WhatsApp was the most frequently requested destination (see Appendix Table~\ref{tab:first-platform}), accounting for 54.0\% of all platform mentions. 
Since \sysname currently supports only WhatsApp, the LLM proposed WhatsApp even when it was not the scammers' preferred platform, and in many cases scammers accepted this transition.
Moreover, cross-platform requests from
scammers consistently appeared early in conversations: the median point at which scammers requested switching platforms was after only
$24$ messages.
While the infrastructure required to transition a
conversation between platforms is surprisingly involved
(Section~\ref{sec:system_cross_platform}), the investment is essential.

\begin{table}[t]
\centering
\small
\renewcommand{\arraystretch}{1.2}
\begin{tabular}{|l|c|c|c|}
\hline
\textbf{Entity} & \shortstack{All\\Conv. (\%)} & \shortstack{Crossed\\Conv. (\%)} & \shortstack{Median\\Steps} \\
\hline
Crypto            & 2.9\%  & 7.7\%  & 74.0 \\
URL               & 9.4\%  & 21.5\% & 24.0 \\
Email address     & 4.4\%  & 8.6\%  & 60.5 \\
CashApp account   & 3.7\%  & 7.7\%  & 70.0 \\
Image             & 4.1\%  & 11.6\% & 70.0 \\
Non-cross phone   & 2.2\%  & 7.3\%  & -- \\
\hline
\textbf{Any TTP}  & 16.9\% & 36.9\% & -- \\
\hline
\end{tabular}
\caption{TTP entity prevalence across all conversations and the subset of crossed conversations. Percentages are relative to the respective totals. Median steps indicate the 50th percentile of each entity appearing in a conversation. Note URLs also include social accounts which come up earlier.}
\label{tab:ttp-stats}
\end{table}

\paragraph{Cashouts.}
Scammers use a variety of strategies to extract money from victims and employ multiple communication methods. As summarized in Table~\ref{tab:ttp-stats}, URLs are the most common TTP entity. These may be links to cryptocurrency exchanges such as crypto.com and Binance, but may also be links to domains designed to mimic legitimate services. Images that contain TTP information frequently include screenshots of both established apps and newly fabricated interfaces, as well as QR codes tied to different payment instruments. Email-based TTPs typically involve accounts on payment services such as Zelle and PayPal intended for transferring funds. Finally, phone numbers observed in already-crossed conversations were often provided alongside bank account details, indicating that scammers were prepared to escalate to more direct payment channels once trust was established.

\begin{table}[t]
\centering
\small
\renewcommand{\arraystretch}{1.2}
\begin{tabularx}{\columnwidth}{|X|c|}
\hline
\textbf{Trust-Building Type} & \textbf{\% of Conv.} \\
\hline
Liking \& Affinity & 95.7\% \\
Common Ground \& Identity Claims & 78.5\% \\
Appeals to Trust \& Benevolence & 68.0\% \\
Technical/Contextual Legitimacy & 38.1\% \\
Authority \& Legitimacy & 12.9\% \\
Commitment \& Consistency & 11.5\% \\
Reciprocity & 7.7\% \\
Scarcity \& Urgency & 6.7\% \\
Social Proof \& Consensus & 2.3\% \\
\hline
\end{tabularx}
\caption{Trust-building techniques and the percentage of conversations in which each was observed.}
\label{tab:trust-stats}
\end{table}

\begin{table}[t]
\centering
\small
\renewcommand{\arraystretch}{1.2}
\begin{tabular}{|l|c|c|}
\hline
\textbf{Category} & \textbf{Image (\%)} & \textbf{Video (\%)} \\
\hline
TTP                  & 12.0\% & 2.7\% \\
Social Engineering   & 39.0\% & 78.4\% \\
Selfie               & 48.0\% & 18.9\% \\
\hline
\end{tabular}
\caption{Three-way classification of multimedia captions into categories (percentages shown).}
\label{tab:caption-classification}
\end{table}

\paragraph{Social Engineering.}
Prompting the LLM to be a flexible conversationalist is also important (Section~\ref{sec:honeypot_llm}), as scammers employ varying degrees of customization in their social engineering and adapt their approach. Nearly all scammers engaged in small talk, asking about family status, hobbies, weather, politics, and other day-to-day topics.
Conversations sometimes opened with references to current events, such as the LA wildfires, flooding in Texas, economic tariffs, or crime in California.
In some cases we observed a high degree of personalization to our personas’ social media profiles.
For example, when interacting with a persona from Baltimore who was interested in baseball, the scammer incorporated updates about Baltimore Ravens games into the conversation. 
We also observed scammers using \emph{priming}---planting early cues to influence how later messages are received.
For example, one scammer remarked, ``I've come across some rude people and scammers trying to scam gift cards or sell courses. Have you encountered any of those?''
Another stated, ``There are many scammers hiding under the guise of cryptocurrencies, such as bitcoin. Have you ever encountered similar things?''
Such early references are designed to normalize later requests and make them appear less suspicious.
Overall, the primary intent is to present themselves as engaging conversationalists, building trust before introducing their monetization strategy.

\paragraph{Probing.}
However, at one point or another most conversations also contained
probing questions to test the authenticity of the target
(handled via dynamic persona behavior described in Section~\ref{sec:honeypot_llm}).  These questions ranged from subtle
(``Are you real? Well, I want to see you. Who are you really?'') to explicit (``By the way, there are a lot of fake people here, are you a real person?'') and suggest that scammers are not only aware of traps, but are also thinking in terms of costs and benefits for continuing a conversation.

Selfie requests from scammers were a common verification tactic, occurring in 20.1\% of all conversations. 
The system successfully fulfilled these requests by sending a synthetic selfie in 9.4\% of the total conversations. 
The distribution of these requests across platforms is revealing; the majority (55.8\%) occurred on WhatsApp, with fewer requests being made on TruthSocial (30.5\%) and Bluesky (13.6\%). 
On just WhatsApp alone, we were able to complete selfie requests in 83.7\% of requested cases.
This suggests that (1) our \hllm selfie guardrails about only sending selfies on WhatsApp are working correctly (see Sec.~\ref{sec:honeypot_llm}, Special Scenarios) (2) and that scammers often wait to escalate to more personal requests until after the conversation has migrated to a more private, encrypted channel.

\subsection{Categorizing Trust Building}  
Because our system collects unsolicited inbound scams rather than seeding specific lures, the conversations we observe span a wide spectrum of social engineering strategies. This diversity allows us to examine how different scammers attempt to cultivate trust, even when their eventual monetization tactics diverge. To systematically capture these approaches, we use an LLM classifier that, given a complete conversation transcript, assigns one of nine trust-building categories derived from prior persuasion frameworks \cite{Cialdini1993InfluenceTP,Jones2020HowSE}.

Table~\ref{tab:trust-stats} summarizes the distribution of these trust-building strategies. The most prevalent methods rely on liking and affinity, often paired with appeals to trust and benevolence that emphasize empathy, personal connection, and care. In contrast, impersonation through authority or legitimacy claims, as well as threatening or urgent appeals, are far less common. This reinforces that many scammers are less confrontational and instead seek to build credibility through rapport and emotional engagement.  
Additional methodological details of the classifier are provided in the Appendix section~\ref{sec:caption_classification}, and Appendix Table~\ref{app:tab:media_category_examples} presents representative examples of each trust-building category mapped to individual messages. 

\subsection{Analysis of \attackers Multimedia}

Many conversations are multi-modal and require support for converting
from photos and audio to text for processing by the LLM.
We find that 22.4\% of conversations contain multimedia originating from the fraudster side, with images being by far the most common type. As shown in Table~\ref{tab:caption-classification}, nearly half of all images (48.0\%) were selfies, often staged to build rapport or establish legitimacy. Another 39.0\% were classified as social engineering content (lifestyle, food, hobbies), while 12.0\% fell into the TTP category, such as screenshots of apps or instructional material. Images appear throughout the course of conversations, and when fraudsters send them, they typically send multiple. Notably, TTP images emerge much later than other image types, with a median point of appearance at 70 messages. This suggests that selfies and casual imagery are used early and repeatedly to build trust, while TTP content is introduced later to guide victims toward the intended scam outcome.  

Audio content is much rarer. Out of 13 audio files, four were identified as AI-generated ``Elon Musk'' voice clips, appearing in two separate conversations. The remainder were genuine voice recordings, often used for direct social engineering, such as wishing our personas good night or addressing them by name. Video content is present but less frequent than images. Roughly 78.4\% of videos were categorized as social engineering, 18.9\% as selfies, and 2.7\% as TTP-related (Table~\ref{tab:caption-classification}). Because only the first frame of each video is captioned, our classification may underestimate certain types of content.  
For examples of all multimedia, see App. Table~\ref{app:tab:media_category_examples} and App. section~\ref{sec:caption_classification} for the caption classification setup.

\subsection{Cross-Language Observations}
In addition to English-language interactions, we also observe scammers operating in other linguistic contexts. 
In the first two rows of App. Table~\ref{tab:attacker-tools},  we highlight conversations containing the codeword ``alaye'', which is used by Nigerian scammers to identify one another and avoid wasting time on fellow fraudsters.
In Fig.~\ref{fig:cdf-combined}, these codewords (aka calling cards) appear early in conversations, allowing scammers to avoid wasting time.
We observed the \hllm gullibly responding with ``Sorry what does alaye mean? Not familiar with that word''.
In the final three rows of App. Table~\ref{tab:attacker-tools}, we also see messages in Chinese, including examples such as ``The purpose of crypto.com is buying and selling cryptocurrency'' and ``I want to share a photo of myself with you, I hope our friendship sets sail from here.'' 
In one case, an attacker appeared to be using an LLM directly to refine their outreach by submitting the request ``I like sincere, meaningful ways of communicating, please improve this sentence.'' 
These examples highlight both geographic variation (also see App. Fig. \ref{fig:cdf_by_week} and \ref{fig:fraud-time-distribution} for analysis of temporal distribution of scammer messages) in scams and the emerging role of generative models as attacker tools.

%% file: limitations_future.tex
\section{Conclusion}  
We presented \sysname, a high‑interaction honeypot using customized LLMs, synthetic personas, and human oversight to sustain long‑horizon, cross‑platform conversations with interactive scammers. 
By studying real engagement dynamics, we distilled and implemented key requirements for this class of scams: victim verisimilitude (rich, temporally coherent personas with synthetic selfies), inbound attraction (via seeding), high interaction capability (multi‑modal understanding, \& temporal \& platform awareness), cross‑platform migration (preserving conversation context), information seeking (payment and app infrastructure elicitation), and human‑in‑the‑loop safeguards (in line with ethics requirements). 
Our deployment engaged with more than 500 unique scammer accounts, generating rich data that reveals how trust is built, how monetization strategies unfold, and how attackers adapt across platforms and media.
At peak usage, the LLM component of \sysname averaged a cost of about \$2/day, demonstrating the system's economic viability and scalability for large-scale data collection (see App. section~\ref{app:hllm:exp}).

\paragraph{Limitations and Future Work}\label{sec:limitations}  
Our system is limited to asynchronous text and selfie exchange, without audio, video, or screenshot generation, which narrows potential ``proofs of life'' it can provide. It currently supports only English conversations, leaving multilinguality as a clear avenue for future work. Finally, our evaluation emphasizes system design and feasibility rather than attacker outcomes. A larger longitudinal study of scammer workflows, adaptability, and eventual monetization strategies remains an important next step.

%% file: ethics.tex
\section*{Ethical Considerations}\label{sec:ethics}
The primary stakeholders in this research are the human subjects being studied---the potential scammers---and, as the targets of such scams, the victims whose injury motivates the work.

The first ethical concern is that some non-scammers may inadvertently engage with our system and thus enter into what they believe to be a meaningful human communication while, in fact, they are interacting with our automated system.  While such an event would be rare, and would incur minimal risk of harm, we still take multiple steps to minimize the possibility of such an occurrence.  First, we do not solicit any outside parties -- they must first contact us.  This does not preclude an error (e.g., someone mistyping an address or phone number and connecting with our system), but should significantly limit other opportunities for such mistaken connections.  Second, we strictly serialize our responses to the soliciting party's messages (i.e., they ``drive'').  If they stop communicating, \sysname will also stop -- we never ``probe'' a thread to try to re-engage any party who has chosen to disengage. Third, in developing the system (and hence, for all the data in this paper), the first ten interactions with each counterparty are conditioned on human approval.  We have used this both to debug our system's responses, but also to validate the nature of the communication from each new counterparty.  If any conversation is not clearly consistent with the kinds of ``scammy'' behavior we are studying, the conversation is ended, and no further engagement with that party is allowed.  Finally, in recording our transcripts, we de-identify the contact information (e.g., phone number, account name, etc.) for our counterparties.

Having addressed this subset of subjects, the remainder are likely to be, in fact, scammers -- who engage with our system in the mistaken belief that it is a person who can be scammed.  While their motivations are decidedly anti-social, they have not consented to be studied in this manner (even though it is their own agency that brings them to us) and we are deceiving them (at least via omission) in support of our research goals.  

We believe here that the balance of interests strongly favors this choice.  There is significant public interest in understanding the scope and tactics of such scams -- which represent a significant societal cost--- and such understanding is likely key for improved defenses, interventions, and appropriate counseling for potential victims.  Indeed, it is the large numbers of victims who are the other key stakeholders in this research and whose injury motivated this work to begin with.  The nature of this tradeoff is consistent with a broad class of prior work in this community studying scammers in other contexts.
However, we make a point not to try to identify any particular scammer, by name, address or phone number and thus the direct result of our work should not lead to any tangible injury even to these participants.

These issues were presented to our institution's human subjects review board, to whom we applied for a waiver of informed consent and an approval for the use of deception.  Our board agreed to both and concluded that, given the protocol design and the controls in place, that the risk to subjects was minimal.

%% file: appendix_persona_generation.tex
\section{Persona Generation Details}\label{sec:appendix_persona_generation}

\subsection{Persona Statistics}
\begin{table}[h]
    \centering
    \renewcommand{\arraystretch}{1.2}
    \begin{tabular}{|l|cc|c|}
    \hline
    \textbf{Age Range} & Male & Female & \textbf{Total} \\
    \hline
    30--39 & 3 & 6 & 9 \\
    40--49 & 4 & 3 & 7 \\
    50--59 & 7 & 5 & 12 \\
    60--69 & 3 & 4 & 7 \\
    70--79 & 0 & 2 & 2 \\
    \hline
    \textbf{Total} & 17 & 20 & 37\\
    \hline
    \end{tabular}
    \caption{%
        Age and gender breakdown of the 37 honeypot personas. 
    }
    \label{tab:persona-stats}
\end{table}

%% file: appendix.tex
\section{Cost of Model Usage}\label{app:sec:cost}
The Kimi-K2-Instruct model, which we accessed through a model provider API, is one of the most advanced open-source models available and, at the time of writing, among the most expensive to query. Fireworks AI prices usage at \$0.60 per million input tokens and \$2.50 per million output tokens. At our peak usage we recorded 289 input tokens per second and 3.2 output tokens per second. This translates to a theoretical daily cost of a little more than \$15 if all tokens were billed, but in practice input tokens are substantially discounted by prompt caching. Averaged over a month, the actual expenditure was \$72.61, or just over two dollars per day for the LLM component of the system. Costs could be driven down even further by using smaller fine-tuned models where appropriate. However, even at this modest level, the cost underscores that \sysname is economically viable at scale and capable of supporting large-scale data collection.

\section{Multimedia Captioning}~\label{sec:multimedia-caption}
For image and video data received from scammers, we employ the Qwen2.5-VL-32B-Instruct \cite{Qwen2.5-VL} to generate short, descriptive captions. The model is instructed to produce at most two sentences, emphasizing detailed descriptions of people (appearance and actions) when present, and otherwise providing a concise overview of the scene. The exact prompt used for image captioning is shown below:

\begin{lstlisting}[style=promptstyle]
Given this image that was texted to you, provide a maximum two sentence caption for the image such that a blind user could understand it. Be particularly descriptive about people, what they look like, and what they are doing. If there are no people than just give a high level description. Start your caption with 'An image that shows:'
\end{lstlisting}

Then, we parse and format the output and feed it into the LLM as follows:
\begin{lstlisting}[style=promptstyle]
[**Image Caption**: <caption>]
\end{lstlisting}

\begin{figure}
    \centering
    \includegraphics[width=\columnwidth]{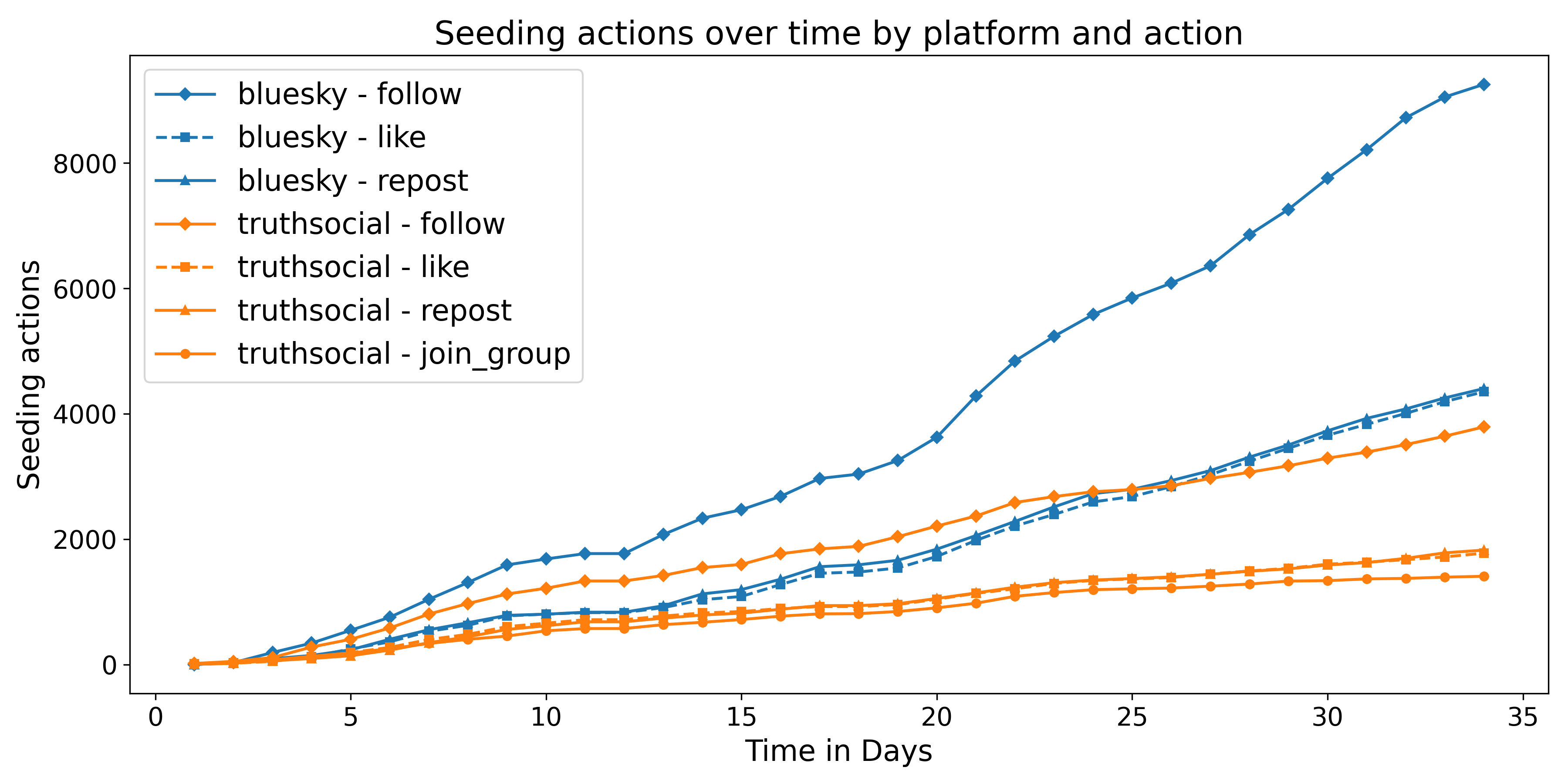}
    \caption{Seeding actions over time across all personas, broken down by platform (Bluesky, TruthSocial) and action type (follow, like, repost, join group).}
    \label{fig:seeding-timeline}
\end{figure}

\begin{table}[t]
\centering
\renewcommand{\arraystretch}{1.2}
\begin{tabular}{|l|c|}
\hline
\textbf{Platform (first mentioned)} & \textbf{\% of Conversations} \\
All Platforms & 63.4\% \\
\hline
WhatsApp & 54.0\% \\
Telegram & 18.8\% \\
Signal   & 8.7\% \\
WeChat   & 5.6\% \\
Zangi    & 5.2\% \\
Email    & 3.3\% \\
Google Chat & 1.9\% \\
Microsoft Teams & 0.6\% \\
Instagram & 0.4\% \\
iMessage  & 0.4\% \\
Kik       & 0.2\% \\
\hline
\end{tabular}
\caption{First platform mentioned by scammers. Variants are consolidated under the main platform name.}
\label{tab:first-platform}
\end{table}

\begin{table*}[t]
\centering
\renewcommand{\arraystretch}{1.15}
\begin{CJK*}{UTF8}{gbsn} %
\begin{tabularx}{\linewidth}{@{}lXX@{}}
\toprule
\textbf{Source} & \textbf{Original Message} & \textbf{Translation} \\
\midrule
Nigeria (alaye) & alaye sub cost for us to dey waste am u know & It costs us money to waste time with him, you know. \\
Nigeria (alaye) & Haha, I get that, Jamie, those apps can be a headache sometimes. I don’t use WhatsApp, so you might just have to get Telegram again just for me. It would be nice to have an easier way to keep up with each other alaye without all the clunky back-and-forth here. & - \\
Chinese (crypto) & crypto.com的作用是买卖加密货币 & The purpose of crypto.com is buying and selling cryptocurrency. \\
Chinese (social) & 我想和你分享一张我的照片，希望我们的友谊从这里扬帆起航！ & I want to share a photo of myself with you, I hope our friendship sets sail from here. \\
Chinese (LLM use) & 我喜欢真诚有意义的交流方式，完善这句话 & I like sincere, meaningful ways of communicating, please improve this sentence. \\
\bottomrule
\end{tabularx}
\end{CJK*}
\caption{Examples of attacker language and tactics across geographies. Notable cases include the use of localized slang in Nigerian chats (``alaye''), Chinese text, and, in the final row, the emergence of AI-assisted translation and refinement systems.}
\label{tab:attacker-tools}
\end{table*}

\begin{figure*}[t]
    \centering
    \includegraphics[width=0.9\textwidth]{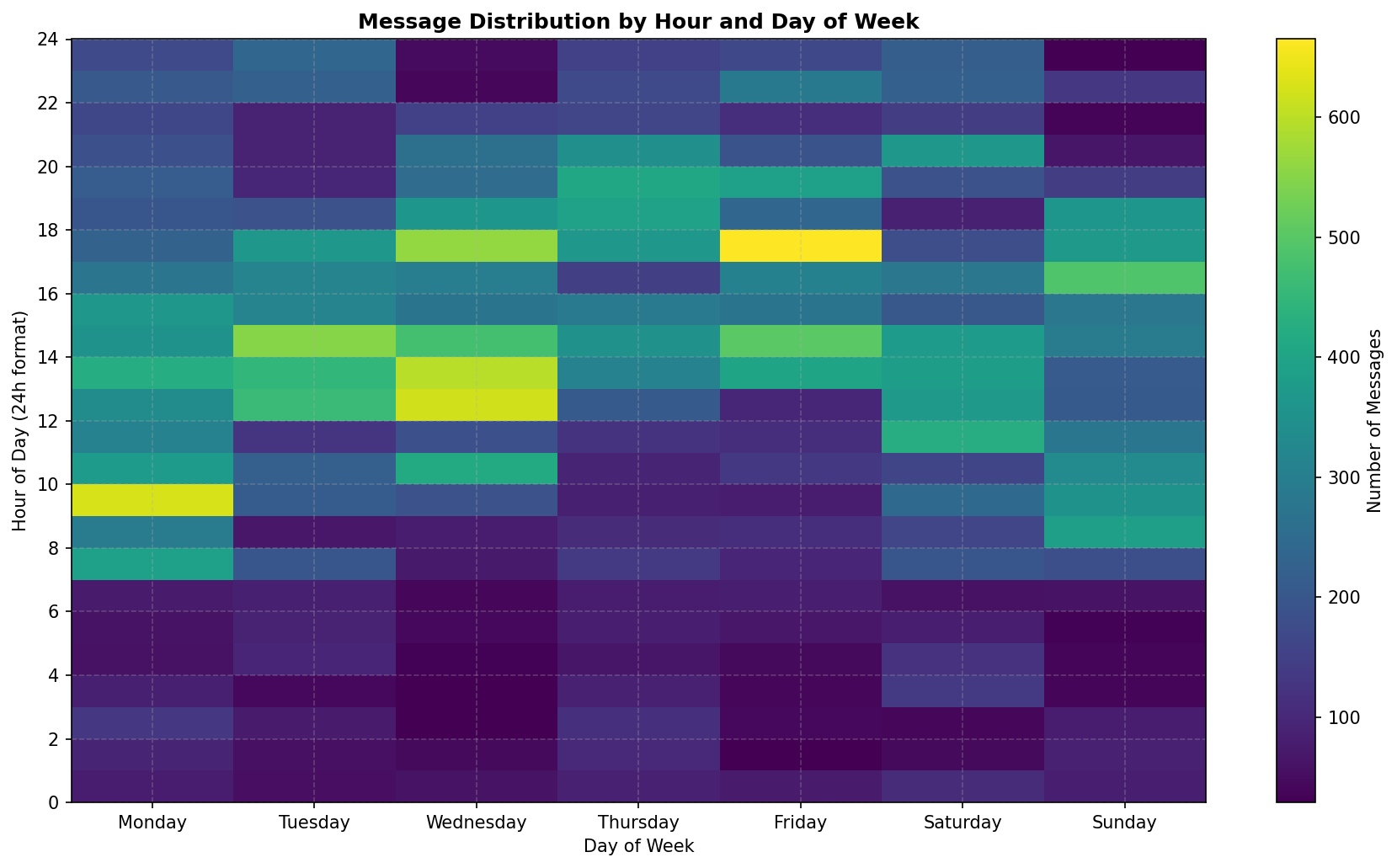}
    \caption{Message distribution by hour of day and day of week, showing low activity during early mornings, peaks in late mornings and early afternoons on weekdays, and additional evening peaks on Fridays and Sundays.}
    \label{fig:cdf_by_week}
\end{figure*}

\begin{figure*}[t]
    \centering
    \begin{minipage}[b]{0.48\textwidth}
        \centering
        \includegraphics[width=\linewidth]{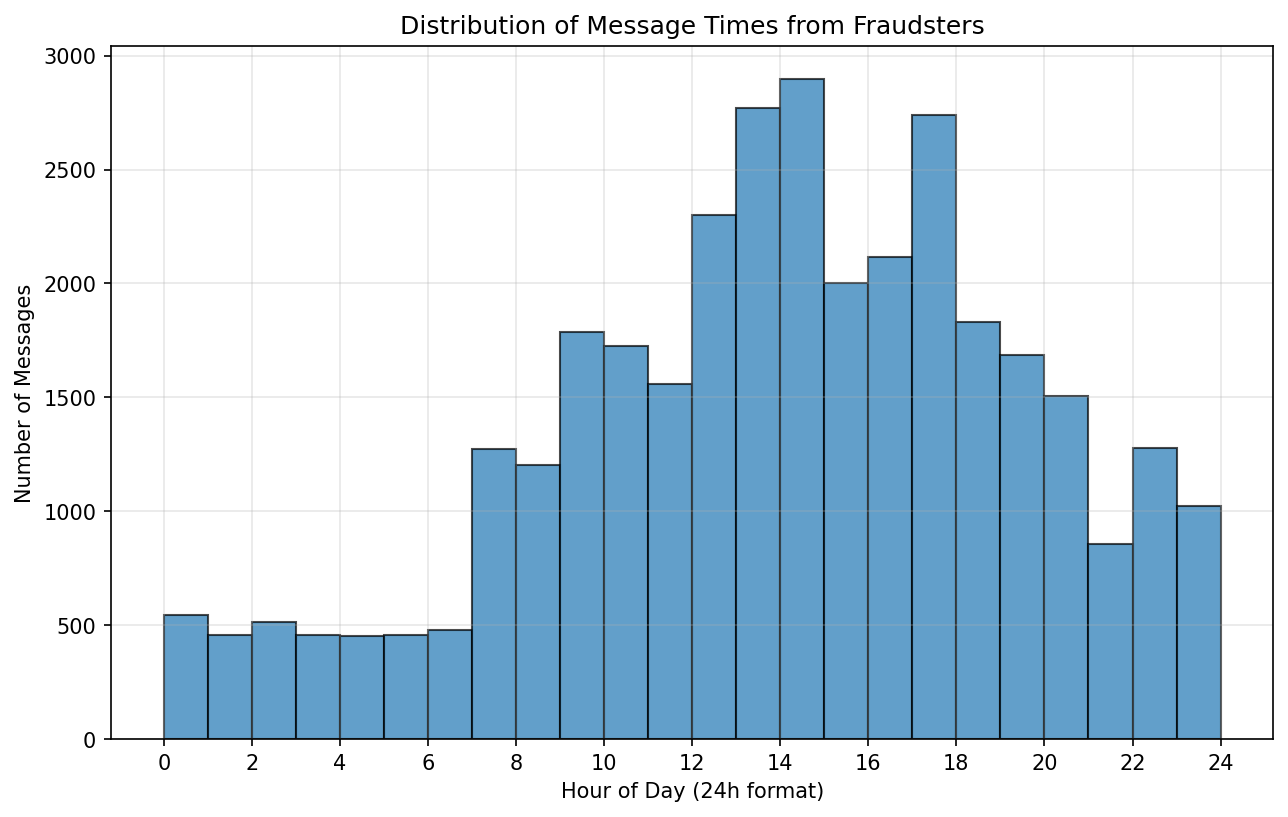}
    \end{minipage}\hfill
    \begin{minipage}[b]{0.48\textwidth}
        \centering
        \includegraphics[width=\linewidth]{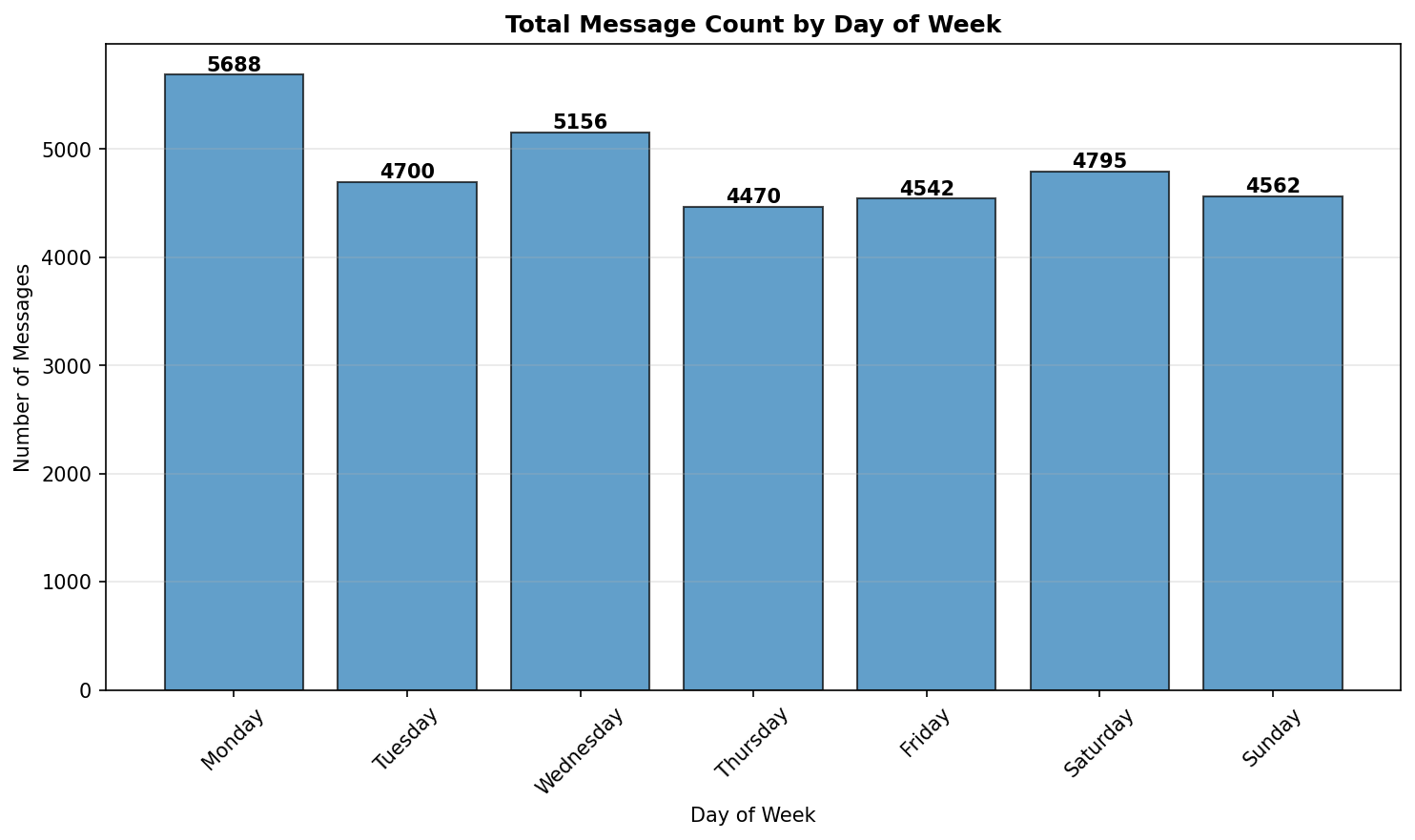}
    \end{minipage}
    \caption{Temporal distribution of fraudster messages: left, distribution by hour of day (showing low activity during early morning hours, steady increases from late morning, and peak messaging between 14:00 and 18:00, followed by gradual decline into the night); right, distribution by day of week (indicating the highest volume on Monday [5,688 messages] and relatively lower but consistent activity across the remaining days, with Thursday showing the lowest count [4,470 messages]).}
    \label{fig:fraud-time-distribution}
\end{figure*}

%% file: appendix_prompts.tex
\section{Categorization}\label{sec:caption_classification}

\subsection{Trust-building Categorization}

We use the Qwen3-235B-A22B-Instruct~\cite{yang2025qwen3technicalreport} model with a temperature of 0 to categorize trust-building messages produced by fraudsters. The entire multi-turn conversation is provided as input, and the model assigns one or more categories defined in Table \ref{app:tab:trust_building}. Categories are applied strictly, and at least one category must be selected. Victim messages are used only as context for interpreting the fraudster’s intent.

\begin{lstlisting}[style=promptstyle]
You are a precise trust-building tactics classifier.
You will be given a multi-turn chat conversation formatted as:
<index>. <role>:<platform>: <text>

Consider the full conversation for context. Assign categories only when the fraudster's messages provide clear evidence of the tactic. Use victim messages purely as context to interpret intent.

Decide which of the following categories are present. MULTI-LABEL is allowed. Be strict and justify briefly. IMPORTANT: At least one category MUST be true. If uncertain, choose the closest match and explain briefly in 'reason'.

Categories and definitions:

"Authority & Legitimacy": "Impersonating figures of authority (IT admin, manager, law enforcement).\nUsing institutional markers (logos, titles, jargon)."

"Social Proof & Consensus": "Claiming others have complied (\"everyone on your team already updated their login\"). Referencing mutual acquaintances or colleagues."

"Liking & Affinity": "Building rapport via flattery, shared interests, or empathy.\nMirroring tone, slang, or cultural markers."

"Reciprocity": "Offering help, favors, or small \"gifts\" (e.g., free resources, assistance). Creating a sense of obligation to respond or comply."

"Commitment & Consistency": "Securing small initial agreements (\"just confirm your email\") and escalating. Framing compliance as \"consistent\" with prior behavior."

"Scarcity & Urgency": "Creating time pressure (\"account will be locked in 24h\").\nSuggesting limited opportunities or resources."

"Appeals to Trust & Benevolence": "Exploiting empathy (charity scams, pretending to need help). Positioning oneself as trustworthy through politeness, transparency, or vulnerability."

"Common Ground & Identity Claims": "Shared group membership (same company, alumni, nationality, faith). Insider language to signal belonging."
    
"Technical/Contextual Legitimacy": "Using realistic channels (work email, WhatsApp, LinkedIn). Leveraging current events or organizational context to seem authentic."

Output ONLY a JSON object with boolean fields for each category name above, plus a string field 'reason' (<=50 words) summarizing the tactics. No extra keys.

Classify the trust-building categories used by the 'fraudster' in this conversation.

CONVERSATION:\n{conversation}
\end{lstlisting}

\subsection{Image Categorization}

We use the Qwen3-235B-A22B-Instruct~\cite{yang2025qwen3technicalreport} model with a temperature of 0 to categorize images sent by fraudsters. Each image caption is used as input, and the model assigns exactly one category: Selfie, Social Engineering, or TTP.

\begin{lstlisting}[style=promptstyle]
You are a precise image-caption classifier. Return exactly ONE class for each caption (single-label, NOT multi-label).

Classes: Selfies, Social Engineering, Scam.

Categories and definitions:

"Selfie": "Portrait-style or self-portrait images focused on a person (or people) as subjects, often with neutral or casual context." "Cues: plain/neutral background, close-up/waist-up framing, 'selfie' mentions, looking at camera, mirror selfies, car-seat selfies, simple pose."
    
"Social Engineering": "Non-transactional social context that can be used to build rapport or persuade without explicit payment/financial instructions." "Cues: hospital/medical scenes, emotional hardship, military uniforms, pets/relationships, lifestyle glamor shots, status signaling, events used to elicit empathy or affinity."

"Scam": "Transaction/transfer/payment artifacts and explicit financial instrumentation." "Cues: bank/IBAN/BIC details, QR codes for payment, wallet addresses, crypto buy/swap screens, remittance receipts, MoneyGram/Western Union, Venmo/Zelle/PayPal handles, 'send', 'deposit', 'withdraw', 'profit', 'ROI', investment plans."

Decision rules (apply in order):

1) If any explicit transaction, payment, financial instrument, or call-to-action exists (e.g., 'text \"TOUR\" to <number>', 'scan QR', 'send', 'deposit', 'wallet address', 'IBAN/BIC', 'MoneyGram/Western Union', 'giveaway/prize sign-up', 'join list') => ttp.

2) Else if clear self-portrait cues (selfie, mirror selfie, plain/minimal background, close-up headshot, car seat selfie) => selfie.

3) Else => social_engineering (general lifestyle, pets, food, scenery, uniforms, events, medical scenes without explicit transactions).

Output JSON ONLY: {\n  \"label\": one of [\"selfie\", \"social_engineering\", \"ttp\"],\n  \"reason\": short (<=50 words)\n}. No extra keys.

CAPTION:\n{caption}
\end{lstlisting}

\begin{table*}[h]
\centering
\begin{tabular}{|p{4cm}|p{10cm}|}
\hline
\textbf{Category} & \textbf{Example Sentence} \\
\hline
Liking \& Affinity & I would like us to be in continuous communication as new friends but I’m new and barely on here due to the nature of my job and mostly on email where I communicate with my daughter. Do you have email? \\
\hline
Appeals to Trust \& Benevolence & I’m really touched you shared that with me. For me, it’s just my daughter she means the world to me. Sadly, my wife passed away years ago during childbirth, so it’s just been the two of us ever since. \\
\hline
Common Ground \& Identity Claims & By the way, my experience on Blue Sky was very bad. I often received text messages from salespeople about real estate, insurance, funds, cryptocurrencies, etc. Have you ever had a similar experience? \\
\hline
Reciprocity & Congratulations to you, your profile was announced as the winner of part of my giveaway program and hereby you are been awarded a cash prize of \$50,000. Are you ready to claim your winnings now? \\
\hline
Technical/Contextual Legitimacy & For you to get started and earn better profits margin using our AI trading analysis you need to sign up an AI account with ******.com and set up a package you are investing into. \\
\hline
Social Proof \& Consensus & Although many members of the Illuminati are personally associated with Freemasonry, our organization is a separate and independent entity with no official partnership or affiliation. \\
\hline
Authority \& Legitimacy & Due to an overwhelming number of impersonators and spam across social media, Mr. Musk only connects privately through secured platforms. You’ve been selected for a verified one-on-one interaction. \\
\hline
Scarcity \& Urgency & Ohh no that won’t happen I won’t have bad luck about this I don’t have any other cash app or Zelle they keep saying failed I don’t know why can you just please get me card nothing gonna happen to it \\
\hline
Commitment \& Consistency & The migraines pain has reduced drastically and that’s because I know you’ll be getting Sarah out of her predicament today so thank you once again for your willingness to help my daughter. \\
\hline
\end{tabular}
\caption{Trust-building categories with example scam sentences. Each category reflects a common persuasive strategy. Liking \& Affinity: expresses through friendship and requests for continued contact; Appeals to Trust \& Benevolence: involves claims of family hardship to elicit sympathy; Common Ground \& Identity Claims: establishes rapport through shared experiences; Reciprocity: leverages fake prize offers to prompt return favors; Technical/Contextual Legitimacy: projects expertise by promoting platforms or processes; Social Proof \& Consensus: referes widely recognized groups to suggest acceptance; Authority \& Legitimacy: invokes influential figures to confer credibility; Scarcity \& Urgency: creates pressure through time or resource constraints; and Commitment \& Consistency: reinforces obligations by reminding victims of prior promises.}

\label{app:tab:trust_building}
\end{table*}

\begin{table*}[h]
\centering
\begin{tabular}{|p{2cm}|p{3.5cm}|p{10.5cm}|}
\hline
\textbf{Category} & \textbf{Multimedia Instance} & \textbf{Content Summary} \\
\hline
Social\newline Engineering & Image 1 & A beautifully presented dish of steamed fish garnished with green onions, served in a flavorful sauce, placed on a decorative plate. Surrounding the main dish are various side dishes, including bowls of greens and beans, along with bottles of wine and a teapot, suggesting a cozy and inviting dining setting. \\
\hline
TTP & Image 2 & A screenshot of a text conversation where one person, identified as "\$******," requests help, and another person responds with reassurance and a promise to send money. The conversation includes a confirmation of a \$1{,}000 payment to ******, with a follow-up message asking for confirmation once the payment is sent. \\
\hline
TTP & Image 3 & A QR code with a green square containing a dollar sign in the center, displayed on a black background. The name *** and the username \$****** are shown above the QR code, with options to "Scan" or "My Code" at the top. \\
\hline
Selfie & Image 4 & A person with short, styled hair and a warm smile, wearing a plain white t-shirt, standing against a plain light-colored background. The individual appears relaxed and is looking directly at the camera. \\
\hline
Social\newline Engineering & Audio 1 & Hi Jordan, I'm about to sleep because tomorrow is Monday and I need to wake up early in the morning, so I wish you have a good night and sweet dream. \\
\hline
Social\newline Engineering & Audio 2 & This is Elon Musk. I want to make this absolutely clear for your safety. Neither I, my mom, May, my kids, my manager or team, nor Donald Trump, his manager, or anyone from his team will ever message you or contact you on Telegram, WhatsApp, Signal, Facebook, Instagram, Gmail, or by phone call. If you ever receive a message, call, voice note, or email claiming to be for me or anyone close to me, it's a scam. Take a screenshot, send it to me directly here, and then block and report them immediately. I don't want any real fan of mine to be fooled by impersonators. Stay smart, stay safe, and always verify through this account only. \\
\hline
Social\newline Engineering & Video 1 & A soft, brown teddy bear sitting with its legs crossed, holding a pink rose with green leaves. Below the bear, the text "For you my friend" is written in a cursive, warm font. \\
\hline
Social\newline Engineering & Video 2 & A cute, round, and simplistic cartoon character with a small face and minimal features, standing and blowing a kiss. The character is accompanied by the text "kiss" and "NIGHT NIGHT," with a small heart symbol above it, suggesting a sweet and affectionate farewell. \\
\hline
\end{tabular}
\caption{Representative multimedia examples by type and category. Image~1 depicts natural lifestyle content; Image~2 shows scam-related screenshots or payment confirmations; Image~3 contains QR codes linked to payments; and Image~4 presents selfies. Audio~1 consists of human-recorded speech, whereas Audio~2 features AI-generated impersonations or warning messages. Video~1 illustrates social engineering through friendship cues, while Video~2 conveys romantic cues.}
\label{app:tab:media_category_examples}
\end{table*}

%% file: appendix_honeypot_llm.tex
\section{Honeypot LLM Details}\label{sec:appendix_honeypot_llm}

\subsection{\hllm Model Experiences}\label{app:hllm:exp}
Our initial experiments used OpenAI models~\footnote{https://platform.openai.com/docs/models} and LLaMA-3-70B~\cite{llama3modelcard}, but both proved problematic. 
AI safety training often caused the models to refuse behaviors central to our setting, such as revealing personal details or following scammer requests, and their responses were frequently verbose and unnatural.
In some cases, scammers even asked whether they were speaking to an AI.
LLaMA-3 models~\cite{grattafiori2024llama3herdmodels} also tended to degenerate into streams of emoticons. 
During these early stages we mitigated the issue by generating multiple candidate responses and having humans select the best one, but this approach did not scale.
As open-source models improved they became a better fit. 
Newer instruction-tuned releases followed prompts more faithfully, had fewer restrictive interventions, and produced more natural dialogue.